\documentclass[prb,twocolumn,showpacs,superscriptaddress,preprintnumbers,amssymb]{revtex4-1}
\usepackage{graphicx}
\usepackage{color}
\usepackage{dcolumn}
\usepackage{bm}
\usepackage{physics}

\newcommand{\beq}{\begin{equation}}
\newcommand{\eeq}{\end{equation}}
\newcommand{\beqn}{\begin{eqnarray}}
\newcommand{\eeqn}{\end{eqnarray}}

\newcommand{\ua}{\uparrow}
\newcommand{\da}{\downarrow}
\newcommand{\ra}{\rightarrow}

\newcommand{\cL}{ {\cal L} }

\newcommand{\cS}{ {\cal S} }

\newcommand{\vect}[1]{{\bm{#1}}}

\newcommand{\ii}{\mathrm{i}}

\newcommand{\SO}{\mathrm{SO}}
\renewcommand{\O}{\mathrm{O}}
\newcommand{\SU}{\mathrm{SU}}
\newcommand{\U}{\mathrm{U}}

\newcommand{\cx}[1]{{\color{black} #1}}

\begin{document}


\title{Pristine and Pseudo-gapped Boundaries of the Deconfined Quantum Critical Points}

\author{Nayan Myerson-Jain}
\affiliation{Department of Physics, University of California, Santa Barbara, CA 93106}

\author{Xiao-Chuan Wu}
\affiliation{Kadanoff Center for Theoretical Physics \& Enrico Fermi Institute, University of Chicago, Chicago, IL 60637}

\author{Cenke Xu}
\affiliation{Department of Physics, University of California, Santa Barbara, CA 93106}

\date{\today}

\begin{abstract}

Bulk topology and criticality can both lead to nontrivial boundary effects. Topological orders are often characterized by their robust edge states, while bulk critical points can have different boundary scalings governed by boundary conditions. The interplay between these two different boundary effects is an intriguing problem. The boundary of the deconfined quantum critical point (DQCP) is the ideal platform for the interplay of the two boundary effects, as the DQCP is also an intrinsically gapless symmetry protected topological (igSPT) state. In this work we discuss the boundary of several analogues of the DQCP. We demonstrate that the fluctuation of the bulk order parameters and their various boundary conditions lead to a rich possibility of the edge states, including a ``pseudogap" (or super power-law decay) behavior. We also discuss the quantum information perspective of our work, i.e. the implication of our results on DQCP under weak-measurement. Weak-measurement followed by post-selection can change the boundary condition at the temporal boundary in the path-integral representation of a density matrix, which will lead to different behaviors of the ``strange correlator".

\end{abstract}

\maketitle

\section{Introduction}

Two distinct classes of systems are both characterized by their boundary effects. Topological states and symmetry protected topological (SPT) states~\cite{wenspt,wenspt2} such as topological insulators usually have robust edge states, which are protected by the 't Hooft anomaly of the system. On the other hand, bulk quantum critical points can also have distinct boundary scalings captured by the boundary conformal field theories (BCFT) \cite{cardybCFT1, cardybCFT2, cardyboundary, LewellenBCCO}. Very recently, a new type of boundary of $3D$ CFTs or $(2+1)d$ quantum critical points (QCP) have been identified, dubbed the ``extraordinary-log" boundary, which further extended the traditional understanding of BCFT~\cite{maxboundary,maxboundary2,toldin1,toldin2,lvboundary}. When these two types of boundary effects happen simultaneously, e.g. a QCP that also involves SPT states, rich and novel possibilities are expected. This interplay has motivated many recent numerical and theoretical works~\cite{zhang1,zhang2,stefan1,stefan2,groveredge,edgexu1,edgexu2, ZhangFQSHExtralog,ThorngrenWanganomalousbdry, Ryuanomalousbdry}.

The DQCP is an unconventional quantum phase transition beyond the classic Landau's paradigm~\cite{deconfine1,deconfine2}. It has been understood that the low energy effective of the DQCP has a 't Hooft anomaly. Therefore, a $(2+1)d$ DQCP can only be realized in
three scenarios: 

(1) It can be realized in a $(2+1)d$ system, when some of the symmetries are not ``on-site". This is the scenario of the original N\'{e}el-VBS DQCP, where the symmetry of the VBS order involves lattice transformation;

(2) All symmetries of the DQCP are on-site and have 't Hooft anomaly, but then the DQCP can only be realized as the $(2+1)d$ boundary of a $(3+1)d$ bosonic SPT state~\cite{senthilashvin,SO5}.

(3) Gapless SPT states is a subject that has attracted great attention and effort~\cite{gspt1,gspt2,gspt3,gspt4}. Most recently it has also been understood that the DQCP can be realized as an ``intrinsically gapless SPT state"~\cite{igspt1,igspt2} in $(2+1)d$, when the low energy symmetry of the system has a 't Hooft anomaly, but it is embedded into a UV symmetry $G$ which is free of anomaly.

Although the DQCP was first proposed as a transition between the N\'{e}el and valence bond solid (VBS) order, it was later realized that the transition between the quantum spin Hall (QSH) insulator and an $s-$wave superconductor (SC) should also be described by the DQCP~\cite{groversenthil,qshscfakher,qshscfakher2,skyrmionsc}. Please note that here the system has a full $\SU(2)\times \U(1)$ symmetry, and the QSH order parameter fluctuates as a vector under SU(2). The $\SU(2)\times \U(1)$ QSH-SC DQCP is an example of intrinsically gapless SPT (igSPT) state, as it involves gapped fermions: the bosonic sector of the system is gapless and has a 't Hooft anomaly; while the fermionic sector is gapped. At the $(1+1)d$ boundary of the $(2+1)d$ igSPT state, i.e. the interface between the DQCP and a trivial state, nontrivial boundary physics must happen in order to neutralize the anomaly of $G_{\mathrm{low}}$.

Hence the boundary of the DQCP is the ideal example of the interplay between the two types of boundary effects mentioned above. 
The current work is partly motivated by the recent numerical observation~\cite{fakhertalk} that, at the boundary of the $\SU(2)\times \U(1)$ DQCP between the QSH phase and the $s-$wave SC, there is a sharp fermion excitation at $\omega = 0$ in the fermion spectral function, or in other words the boundary fermion modes of the QSH insulator appear ``pristine" even at the QSH-SC DQCP. In this work we will show that, a pristine boundary where the fermion Green's function scales similarly to that of a free fermion topological insulator (up to a logarithmic correction) is indeed possible for DQCP, but there also exists another possibility where the fermions acquire a ``pseudogap"; more precisely the fermion Green's function decays as a super power-law with distance.



\begin{table}

\begin{tabular}{lll}

\hline \hline \multicolumn{1}{l|}{QSH$_z$-SC }   & $\delta r > 0$ (eq.~\ref{1d1}):  $\sim 1/|x|$ &  \\ \cline{2-3} \multicolumn{1}{l|}{Transition} & $\delta r < 0$ (eq.~\ref{1d1}): $\frac{\exp( - \frac{2}{q} \log(|x|)^2)}{(\log |x|)^{q/4}} $ &  \\

\hline \hline \multicolumn{1}{l|}{Large-$N$ }   & $\delta r > 0$ (eq.~\ref{1d2}): $\frac{1}{|x|^{2 \Delta_b}} \frac{\exp(- \frac{2}{q} \log(|x|)^2)}{(\log |x|)^{q/4}}  $ &  \\ \cline{2-3} \multicolumn{1}{l|}{Easy-Plane DQCP} &    $\delta r < 0$ (eq.~\ref{1d2}): $\frac{1}{(\log |x|)^{q}} \frac{1}{|x|}$ &

\\ \hline \hline \multicolumn{1}{l|}{$\SU(2) \times \U(1)$}   & $\delta r \ra +\infty $ (eq.~\ref{1d3}): $\sim$ pseudo-gapped  &  \\ \cline{2-3} \multicolumn{1}{l|}{DQCP} & $\delta r \ra - \infty$ (eq.~\ref{1d3}): $\sim$ pristine  &  \\ \hline \hline

\end{tabular}

\caption{Expected spatial decay of the boundary electron Green's functions for each model considered in this work. For simplicity we take the Luttinger parameter $K$ for the boundary fermions (either electron $c_\alpha$ or parton $f_\alpha$) to be $1$. \cx{When the spatial decay of the electron Green's function is close to a free fermion behavior (for example a $1/|x|$ decay with a logarithmic correction), the boundary state is viewed as pristine; on the contrary, if the spatial decay is dominated by a super power-law scaling, it is viewed as ``pseudo-gapped".}} 

\end{table}

However, the boundary of the original $\SU(2)\times \U(1)$ DQCP may not be very well-defined problem to study. Various works have gradually converged to the conclusion that the DQCP is actually a weak first order transition with pseudo-critical behavior~\cite{1storder,loopmodel1,loopmodel2,bootstrap,bootstrap1,SO5,song2024extracting,song2024deconfined,hefuzzy,sandvik1st}.
Therefore in this work we explore the edge state of several related quantum critical points that involve the QSH insulator and the SC order. Some examples that we discuss are analogous to the QSH-SC DQCP studied numerically in Ref.~\onlinecite{fakhertalk}, and in certain limit our example is indeed a continuous DQCP which allows us to extract the information of the edge state. 
We note that a previous theoretical work~\cite{mazouwang} also investigated the boundary of the QSH-SC DQCP with $\SU(2)\times \U(1)$ symmetry, but the analysis therein utilized a theoretical generalization which led to different results from our current work.

\section{QSH$_z$-SC transition}

Let us first consider the simplest related example, which is a quantum phase transition between a $2d$ QSH insulator to a superconductor phase. We can start with a Kane-Mele QSH model~\cite{kane2005a,kane2005b} of fermions on the honeycomb lattice, and couple the fermions to a fluctuating superconductor order parameter $\Phi$ in the bulk. The QSH-SC transition should belong to the ordinary $3D$ XY universality class, described by the following action: \beqn \cS = \int d^2x d\tau \ |\partial_\mu \Phi|^2 + r |\Phi|^2 + u |\Phi|^4, \label{XY}\eeqn with $r$ tuned to zero in the bulk. Note that here the QSH order parameter in the bulk is {\it fixed} along the $z$ direction rather than fluctuating (hence the transition is dubbed the QSH$_z$-SC transition). If we ignore the fluctuations of the SC order parameter $\Phi$, the bulk QSH order will lead to a sharp helical $(1+1)d$ boundary state described by the following action \beqn \cS_{(1+1)d} = \int dx d\tau \ c^\dagger (\partial_\tau - \ii \sigma^z \partial_x ) c + \cL^{\mathrm{int}}. \eeqn Here $c_{\alpha}$ with $\alpha = \ua, \da$ corresponds to two spin-species of electrons, and also the left and right moving electrons. The term $\cL^{\mathrm{int}}$ includes the interactions between the boundary fermion modes, which changes the Luttinger parameter $K$ and drive the boundary to an interacting Luttinger liquid.

Whether the boundary fermionic modes remain a sharp excitation or not, depends on their coupling to the fluctuating SC order parameter $\Phi$ at the $1d$ boundary. Depending on the details of the Hamiltonian at the boundary, the order parameter $\Phi$ can receive an extra boundary mass term, and it will couple to the boundary fermion modes in the following way \beqn \cS_{(1+1)d} = \int dx d\tau \ \delta r |\Phi|^2 + g \Phi^\ast (\epsilon_{\alpha\beta} c_{\alpha} c_{\beta}) + h.c. \label{1d1}\eeqn When $\delta r > 0$, there is a positive extra mass term of the order parameter $\Phi$ at the boundary of the system, which should drive the boundary to the ``ordinary" boundary of $\Phi$. Near the ordinary boundary, $\Phi$ should be replaced by $\partial_y \Phi$, which substantially increases its scaling dimension. Detailed calculation such as an $\epsilon$-expansion can be carried out to evaluate the boundary scaling dimension $\Delta_b[\Phi]$ of $\Phi$, and it is known that $\Delta_b[\Phi] > 1$~\cite{cardybook,cardyboundary,boundary2,boundary3,boundary4,boundary5}. It is obvious that if the boundary fermionic modes do not have self-interaction (i.e. the Luttinger parameter of the boundary Luttinger parameter $K = 1$), the coupling between $c_\alpha$ and $\Phi$ is irrelevant, as $\Delta[g] < 0$. Then unsurprisingly, the edge state of the noninteracting QSH insulator remains pristine when the SC order parameter is in its ordinary boundary phase.

With certain amount of interaction between $c_\ua$ and $c_\da$, the Luttinger parameter of $c_\alpha$ can render the coupling $g$ relevant. This is best captured in the bosonized formalism of the boundary modes $c_\alpha$: \beqn \cS_{(1+1)d} = \int dx d\tau \ \frac{1}{2\pi K} (\partial_\mu \theta)^2 + \tilde{g} \Phi e^{\ii 2\theta} + h.c. \label{}\eeqn The fermion operator $c_{\ua,\da}$ is identified as $c_{\ua,\da} \sim e^{\ii \theta \pm \ii \phi}$, where $\phi$ is the dual boson of $\theta$. The Luttinger parameter $K$ depends on the density-density interaction between the left-moving (spin-up) and right-moving (spin-down) fermions $c_{\ua,\da}$, and when $K < K_c$, $\tilde{g}$ becomes relevant. Based on the analysis in Ref.~\cite{maxboundary}, it is expected that the relevant RG flow of $\tilde{g}$ would lead to an extraordinary-log correlation of $e^{2\ii \theta}$. This phase should be the same as the case with $\delta r < 0$ in Eq.~\ref{1d1}, where $\Phi$ will be in its extraordinary-log boundary phase.

When $\Phi$ has an extraordinary-log boundary, the equal-time Green's function of electrons at the $1d$ boundary will decay with a super power-law \beqn G_c(x) \sim \frac{\exp\left(- \frac{2}{q} (\log |x|)^2\right) }{(\log |x|)^{q/4}} 
\label{green0} \eeqn Here $1/(\log |x|)^{q/4}$ comes from the extraordinary-log correlation of $e^{\ii \theta}$, and $q$ is an exponent associated with the $3D$ XY universality class~\cite{maxboundary}; while $\exp\left(-\frac{2}{q} (\log |x|)^2\right)$ is the correlation of $e^{\ii \phi}$, computed with an effective formalism detailed in the appendix. 
The value of $q$ has been estimated from Monte Carlo results to be roughly between $q \sim 0.5$ and $q \sim 0.6$ \cite{maxboundary,lvboundary, toldin2}.
The fermion correlation decays
faster than any power-law, but slower than exponential, i.e. the boundary fermion modes acquire a ``pseudogap" as expected, because the extraordinary-log boundary of $\Phi$ is an almost long range ordered phase of superconductor.

\section{The parton/dual description}

It is well-known that the $3D$ XY quantum critical point has a dual description in terms of vortices~\cite{peskindual,halperindual,leedual}: \beqn \cS_{\mathrm{dual}} = \int d^2x d\tau \ |(\partial_\mu - \ii a_\mu)\phi_v|^2 + \tilde{r} |\phi_v|^2 + \tilde{u} |\phi_v|^4 + \cdots \label{dual1} \eeqn where $\tilde{r} \sim - r$. The Maxwell term of the dynamical gauge field $a_\mu$ was not written explicitly. The gauge field $a_\mu$ is the dual of the current of $\Phi$ through the relation $J^\Phi_\mu \sim \epsilon_{\mu\nu\rho}\partial_\nu a_\rho$, and $\phi_v$ is the charge of $a_\mu$ which physically corresponds to the vortex of $\Phi$. It is known that the ordered phase of $\Phi$, i.e. the superconductor phase, corresponds to the photon phase of $a_\mu$; while the disordered phase of $\Phi$ ($r > 0$, or $\tilde{r} < 0$) corresponds to the condensate of $\phi_v$, which gaps out the gauge field through the Higgs mechanism.

To facilitate the large$-N$ generalization related to the DQCP that will be discussed in the next section, we introduce a parton description of the QSH$_z$-SC transition: \beqn c_{r,\ua} = b_r f_{r,\ua}, \ \ \ c_{r,\da} = b^\dagger_r f_{r,\da}. \eeqn Note that partons $b$ and $f_\alpha$ are introduced at every site $r$ of the honeycomb lattice. This parton construction has three important $\U(1)$ symmetry and gauge transformations: \beqn \U(1)_s &:& c_\alpha \ra \left( e^{\ii \theta_s \sigma^z} \right)_{\alpha\beta} c_\beta, \ \ b \ra e^{\ii \theta_s } b, \cr\cr \U(1)_e &:& c_\alpha \ra e^{\ii \theta_e } c_\alpha, \ \ f_\alpha \ra e^{\ii \theta_e } f_\alpha, \cr\cr \U(1)_g &:& b \ra e^{- \ii \theta_g} b, \ \ f_{\alpha} \ra \left( e^{\ii \theta_g \sigma^z} \right)_{\alpha\beta} f_{\beta}. \eeqn $b$ is a bosonic rotor carrying the $\U(1)_s$ spin quantum number, as well as the gauge charge. The bosonic rotor $b$ can have both positive and negative density fluctuation, with zero total average density.

To make connection to the QSH-SC transition, we keep the fermionic parton $f_\alpha$ in a ``QSH" insulator mean field state, in the sense that at the mean field level, $f_{\ua/\da}$ forms a Chern insulator with Chern number $\pm 1$. When $b$ is condensed, $f_\alpha$ and $c_\alpha$ are identified, and the system becomes an ordinary QSH insulator; when $b$ is gapped, the system is a QSH insulator of $f_\alpha$ coupled with the gauge field $a_\mu$. The $2\pi$-flux of $a_\mu$ is a ``spin" gauge flux on $f_\alpha$, and it would carry charge$-2$ of $f_\alpha$, which is also the physical charge of $c_\alpha$. Hence the flux of $a_\mu$ is a conserved quantity as the system has the $\U(1)_e$ symmetry. The space-time monopole of $a_\mu$ which is an event that creates the gauge flux of $a_\mu$, is identified as the superconductor order parameter $\Phi$. When the bosonic parton $b$ is gapped, $a_\mu$ will be in its photon phase, meaning that the flux of $a_\mu$ is in its condensate, which is precisely the superconductor phase. Hence we can identify the coarse-grained field of the bosonic parton $b$ as the vortex field $\phi_v$ in Eq.~\ref{dual1}.


The extraordinary-log boundary condition of $\Phi$ is dual to the ordinary boundary of $\phi_v$, which can be seen by the duality relation $\tilde{r} \sim - r$. An extra positive mass term for $|\phi_v|^2$ at the boundary $y = 0$, corresponds to a negative mass term for $|\Phi|^2$ at the boundary, i.e. \beqn \delta \cS_{(1+1)d} = \int dx d\tau \ \delta r |\phi_v|^2 \ra \int dx d\tau - \delta r |\Phi|^2, \label{dualboundary} \eeqn and a negative mass term of $\Phi$ near the $1d$ boundary would drive $\Phi$ into its extraordinary-log boundary phase, as was discussed previously. 

\section{Large$-N$ easy-plane DQCP}

We generalize the single-flavor problem to the case of $N-$flavors of electrons, each flavor with its own spin-up and down. The flavor index can be viewed as orbital degrees of freedom in a material. And we consider the following parton construction: \beqn c_{r, j, \ua} = b_{r, j} f_{r, \ua}, \ \ \ c_{r, j,\da} = b^\dagger_{r, j} f_{r, \da}, \eeqn where $j = 1 \cdots N$. There is a $[\U(1)_s]^N$ flavor symmetry carried by $b_j$, a $\U(1)_e$ symmetry carried by $f_\alpha$, and a $\U(1)_g$ transformation shared by $b_j$ and $f_\alpha$.

Again, we keep $f_\alpha$ in a background quantum spin Hall insulator mean field state. A bulk quantum phase transition can occur between the superconductor phase where $b_j$ are gapped, to a QSH insulator with spontaneous breaking of the $[\U(1)_s]^N$ flavor symmetry. This transition is described by the following easy-plane CP$^{N-1}$ (EPCP$^{N-1}$) field theory (or a bosonic QED$_3$ with $N$ flavors of matter fields): \beqn \cS = \int d^2x d\tau \ \sum_{j = 1}^N |(\partial_\mu - \ii a_\mu)z_j |^2 + \tilde{r} |z_j|^2 + \tilde{u} |z_j|^4 + \cdots \label{epcpn}\eeqn $z_j$ is the coarse-grained field of $b_j$. Eq.~\ref{epcpn} is a large-$N$ generalization of Eq.~\ref{dual1}. When $z_j$ are all gapped, the system is in a superconductor phase for the same reason as the previous section; when $z_j$ condenses, the system spontaneously breaks the $[\U(1)_s]^N$ flavor symmetries to its $\U(1)$ diagonal subgroup (which is ``gauged" by the $\U(1)$ gauge field $a_\mu$). And there are $N-1$ independent gauge invariant order parameters: \beqn \varphi_j \sim z_j^\ast z_{j+1}, \ \ j = 1, \cdots N-1. \eeqn

Let us now discuss the bulk critical point $\tilde{r} = 0$. It is expected that for large enough $N$, the EPCP$^{N-1}$ model has a continuous phase transition, i.e. $\tilde{r} = 0$ is a $(2+1)d$ CFT~\cite{deconfine2}. And in the large-$N$ limit, the gauge field fluctuation is completely suppressed, the transition is essentially just $N$ copies of decoupled $3D$ XY transition. The reason the $N$ copies of $3D$ XY transitions are decoupled, is because the leading symmetry allowed inter-flavor coupling $|z_i|^2|z_j|^2$ is irrelevant at the $3D$ XY$^N$ universality class~\cite{deconfine2}, as the critical exponent $\nu$ of $3D$ XY transition is (slightly) greater than $2/3$.

Now we investigate the boundary of the system. \cx{First of all, at the $(1+1)d$ boundary, the parton $f_\alpha$ has a helical boundary state at the mean field level: \beqn S_{(1+1)d} = \int d\tau dx \ f^\dagger (\partial_\tau - \ii \sigma^z \partial_x) f. \eeqn The boundary fermion modes $f_\alpha$ couples to gauge field in two ways: it not only will couple to the gauge field $a_\mu$ through the ordinary minimal coupling, but also to the monopole field in the same way as $\Phi$ couples to $c_\alpha$ in Eq.~\ref{1d1}: $\Phi \epsilon_{\alpha\beta} f_\alpha f_\beta + h.c.$. Note that the monopole field of the gauge field is the superconductor order parameter due to the spin-Hall effect.} At the $(1+1)d$ boundary, an extra mass term for $z_j$ can be turned on: \beqn \delta \cS_{(1+1)d} = \int dx d\tau \ \delta r \sum_{j = 1}^N |z_j|^2. \label{1d2} \eeqn When $\delta r$ is {\it negative}, at least in the large-$N$ limit $z_j$ itself can be viewed as an XY order parameter as the gauge field fluctuation $a_\mu$ is fully suppressed, and $z_j$ will be in the extraordinary-log boundary phase. In this limit, the real-space Green's function of $c_{j,\alpha}$ at the boundary is 
\beqn G_c(x) \sim G_b(x) G_f(x) \sim \frac{1}{(\log |x|)^{q}} \frac{1}{|x|}. \label{green1} \eeqn $G_b$ and $G_f$ are the correlation functions of the bosonic rotor $b$ and fermionic parton $f$ along the boundary respectively. $G_b$ takes the form of the extraordinary-log boundary phase of the XY order parameter. Eq.~\ref{green1} is not so different from the free fermion Green's function in $(1+1)d$, and \cx{we still view these boundary fermion modes as the ``pristine" boundary state}. 

In the opposite limit $\delta r > 0$, $z_j$ will be at its ordinary boundary phase. Just like the previous case with $N = 1$ ($e.g.$ Eq.~\ref{dualboundary}), we expect the monopole field of gauge field $a_\mu$, which is essentially the SC order parameter $\Phi$, will be in the extraordinary-log boundary phase. The heuristic reason is that, in the $2d$ bulk when $z_j$ is gapped, the gauge field monopole (or superconductor order parameter) will be in its ordered phase; hence a positive $\delta r$ tends to drive $\Phi$ into an ``ordered" phase. Since $\Phi$ cannot really form a long range order at the $1d$ boundary, it will become the extraordinary-log phase. Note that the extraordinary-log boundary for a $\U(1)$ (or XY) order parameter can always exist, \cx{as long as the OPE between the boundary and bulk fields do not have singularity (please refer to the appendix and Ref.~\cite{maxboundary}).} Also, to analyze the extraordinary-log boundary of $\Phi$, we {\it cannot} take the large-$N$ limit of $b_j$, as this limit will fully suppress the dynamics of the gauge field, or the fluctuation of $\Phi$.

In the case with $\delta r > 0$, the fermion Green's function should still be dominated by a super power-law decay, as at the mean field level, the fermionic parton $f_\alpha$ should essentially decay in the similar way as Eq.~\ref{green0}. An approximate form of the spatial decay of the fermion Green's function would be \beqn G_c(x) \sim G_b(x) G_f(x) \sim \frac{1}{|x|^{2\Delta_b}} \frac{\exp\left(- \frac{2}{q} (\log |x|)^2 \right)}{(\log |x|)^{q/4}} \label{green2}. \eeqn The first factor $\frac{1}{|x|^{2\Delta_b}}$ comes from the correlation of $b_j$ which is in its ordinary boundary condition, and $\Delta_b > 1$ is the boundary scaling dimension of $b_j$ at its ordinary boundary. The rest of the correlation corresponds to $G_f$ which should behave similarly as Eq.~\ref{green0}.


\section{The $\SU(2)\times \U(1)$ QSH-SC DQCP}

The intertwinement between the order parameters at the DQCP can be captured by a $(2+1)d$ O(5) nonlinear sigma model with a WZW term~\cite{senthilfisher}. It has a five component unit vector $\vect{n} = (n_1 \cdots n_5)$, where $\vect{n}_A \sim (n_1, n_2, n_3)$ represent the three component QSH vector, and $\vect{n}_{B} \sim (n_4, n_5)$ corresponds to the real and imaginary parts of the $s-$wave superconductor order parameter $\Phi$, i.e. $\vect{n}_{B} = (\mathrm{Re}(\Phi), \mathrm{Im}(\Phi))$. It was proposed~\cite{senthilfisher,SO5}, and shown numerically~\cite{loopmodel1,loopmodel2} that the $2d$ bulk (which is tuned to the DQCP) has an emergent SO(5) symmetry in the infrared, even though the DQCP in the bulk is a very weakly first order transition. At the $1d$ edge, one can turn on the following term which breaks the SO(5) symmetry down to the physical $\SO(3)\times \SO(2)$ symmetry: \beqn \delta \cS_{(1+1)d} = \int dx d\tau \ \delta r \left( (\vect{n}_A)^2 - (\vect{n}_B)^2 \right). \label{1d3}\eeqn The scaling dimension of $(\vect{n}_A)^2 - (\vect{n}_B)^2$, i.e. the tuning parameter of the DQCP, is smaller than $2$ (assuming DQCP is a continuous transition), hence $\delta r$ is still a relevant perturbation even at the $(1+1)d$ boundary. \cx{Note that one can also turn on a SO(5) singlet operator at the boundary, but since the SO(5) singlet has a scaling dimension close to $3$~\cite{hefuzzy}, a SO(5) singlet perturbation will be highly irrelevant at the $(1+1)d$ boundary. Hence we only consider perturbation Eq.~\ref{1d3} at the boundary. }

\cx{Without loss of generality, $\delta r$ may flow to either $\pm \infty$ in the infrared limit. If $\delta r$ can flow to $-\infty$, the QSH vector $\vect{n}_A$ has a strong tendency to ``order" at the boundary, 
and $\vect{n}_A$ should form
an extraordinary-log boundary phase}. 
Since $\vect{n}_A$ is almost ordered near the boundary, we expect the fermion Green's function to be close to the weakly interacting Fermion Green's function, which would lead to a sharp peak at $\omega = 0$ in the fermion spectral function. Also, when $\delta r$ flows to $-\infty$, the superconductor order parameter $\Phi$ is in an ordinary boundary phase whose scaling dimension is greater than 1 near the boundary, hence its coupling to the boundary fermion arising from the nearly ordered QSH vector $\vect{n}_A$ would be irrelevant.

If $\delta r $ flows to $+\infty$, the superconductor order parameter $\Phi$ will be in an extraordinary-log phase near the boundary, while the QSH vector $\vect{n}_A$ will be in an ordinary boundary phase. In this case, a SU(2) parton analysis similar to Ref.~\cite{mazouwang} would indicate that the fermion Green's function decays with a super power-law. Hence the prediction of this theoretical picture is that, by tuning the Hamiltonian at the boundary, there should be a boundary transition that corresponds to changing the sign of $\delta r$ at the boundary, and the sharp feature of the boundary fermion states would vanish across the boundary phase transition.

Here we stress that, in this case a generalization of the $\SU(2)$ parton to large-$N$ flavors of bosons with a $\SU(N)$ symmetry will likely not capture the key desired physics. A large-$N$ rotor field with a SU($N$) symmetry would no longer have an extraordinary-log boundary, and the dynamics of the gauge field (or the SC order parameter $\Phi$) would be completely suppressed. Hence a large-$N$ generalization with $\SU(N)$ symmetry would not capture the physics of both RG fixed points $ \delta r = \pm \infty$.

\section{DQCP under weak-measurement}

Generally speaking, the 't Hooft anomaly of the boundary of a SPT state arises from topology in the space-time bulk. Hence nontrivial physics should occur at both the spatial boundary, as well as the ``temporal boundary". The 't Hooft anomaly at the temporal boundary can be captured by the following ``strange correlator"~\cite{YouXu2013}: \beqn \frac{\langle \Psi | \hat{O}(0) \hat{O}(\vect{x}) | \Omega\rangle }{\langle \Psi |\Omega\rangle}, \ \ \mathrm{or} \ \ \frac{\tr\left( \rho_\Psi \rho_\Omega \hat{O}(0) \hat{O}(\vect{x}) \right)}{\tr\left( \rho_\Psi \rho_\Omega \right)}. \label{strange} \eeqn Here $|\Psi\rangle$ and $|\Omega\rangle$ are the wave functions of the SPT state and a trivial state respectively, and $\rho_{\psi,\Omega}$ are their corresponding density matrix. Both strange correlators are referred to as ``type-I" strange correlator~\cite{sptdecohere,zhang2024strange,ma2024topological}.

The reason the strange correlator captures the 't Hooft anomaly at the temporal boundary is that, in a path-integral formalism, a density matrix can be evaluated as a path-integral in the $d+1$ dimensional Euclidean space-time weighted by the action derived from the Hamiltonian of the system. Then the strange correlator between two different wave functions (or density matrices) can be viewed as the correlation function at the temporal interface between two actions. If the actions belong to two phases with different topological classification, the temporal interface should also capture the 't Hooft anomaly just like the spatial interface. In our problem we would like to choose $|\Psi\rangle$ to be one of the QSH-SC QCPs discussed above, and $|\Omega \rangle$ to be a trivial insulator of fermion $c_\alpha$. Also we choose $\hat{O}$ to be $c_\alpha$, and $\vect{x} = (x, y)$ is the purely
$2d$ spatial coordinate.

Most recently it was also realized that a class of quantum information problems, i.e. the physics of a quantum many-body systems under weak-measurement or finite-depth quantum channel can be mapped to physics on the temporal boundary in the Euclidean space-time path-integral~\cite{altman1}. In particular, in the path-integral formalism, the temporal boundary condition can be tuned by weak-measurement. This picture has been applied to the mixed state density matrix of SPT states and quantum critical points~\cite{altman1,sptdecohere,wfdecohere,altman2,fan2023,jianmeasure2,zoumeasure,cherndecohere,anyondecohere}, and strange correlators have been used as a diagnosis for mixed SPT states~\cite{sptdecohere}.  

In our problem, in order to tune the temporal boundary condition, one can perform weak-measurement on $T(\vect{x}) = (\vect{n}_A)^2 - (\vect{n}_B)^2$, followed by post-selecting either positive or negative measurement outcomes. The weak-measurement followed by post-selection will turn on an extra term on the $(2+0)d$ temporal interface between the DQCP and trivial state density matrix: \beqn \delta \cS_{(2+0)d} = \int d^2x \ \delta r \left( (\vect{n}_A)^2 - (\vect{n}_B)^2 \right). \eeqn The sign of $\delta r$ can be tuned by selecting the positive or negative measurement outcomes of $T(\vect{x})$. Then the strange correlator of $c$ and $c^\dagger$ should behave like either Eq.~\ref{green1} or Eq.~\ref{green2} depending on the sign of the fixed points of $\delta r$.

\section{Summary}

In this work we discussed the boundary physics of several examples of analogues of the $\SU(2)\times \U(1)$ DQCP, when the DQCP is realized as an intrinsically gapless SPT state. We demonstrate that the boundary condition, especially the recently discovered extraordinary-log boundary of the competing order parameters would lead to rich possibilities of the boundary fermion modes. \cx{In particular, we discuss two possibilities of the boundary fermion modes: they can be either ``pristine", or ``pseudo-gapped”, depending on which of the competing orders are in the ordinary or extraordinary-log boundary phases. }

One can also embed the original N\'{e}el-VBS DQCP into a fermion model, such as the interacting Su-Schrieffer-Heeger (SSH) model. In this case, the physical boundary of the system would explicitly break {\it part} of the VBS symmetry, and may lead to even more possibilities of the boundary fermion modes. We leave the investigation of this model to future studies.

The authors thank Fakher Assaad, Chao-Ming Jian, Max A. Metlitski, Francesco Parisen Toldin for helpful discussions. C.X. acknowledges support from the Simons foundation through the
Simons investigator program. X.W. is supported in part by the Simons Collaboration on Ultra-Quantum Matter, which is a grant from the Simons Foundation (651440), and the Simons Investigator award (990660). This research was supported in part by grant NSF PHY-2309135 to the Kavli Institute for Theoretical Physics (KITP). While finishing the current paper, we became aware of another work which numerically studies the boundary of a model similar to our QSH$_z$-SC transition~\cite{toldinfuture}. 

\bibliography{gaplessSPT}

\begin{thebibliography}{71}%
\makeatletter
\providecommand \@ifxundefined [1]{%
 \@ifx{#1\undefined}
}%
\providecommand \@ifnum [1]{%
 \ifnum #1\expandafter \@firstoftwo
 \else \expandafter \@secondoftwo
 \fi
}%
\providecommand \@ifx [1]{%
 \ifx #1\expandafter \@firstoftwo
 \else \expandafter \@secondoftwo
 \fi
}%
\providecommand \natexlab [1]{#1}%
\providecommand \enquote  [1]{``#1''}%
\providecommand \bibnamefont  [1]{#1}%
\providecommand \bibfnamefont [1]{#1}%
\providecommand \citenamefont [1]{#1}%
\providecommand \href@noop [0]{\@secondoftwo}%
\providecommand \href [0]{\begingroup \@sanitize@url \@href}%
\providecommand \@href[1]{\@@startlink{#1}\@@href}%
\providecommand \@@href[1]{\endgroup#1\@@endlink}%
\providecommand \@sanitize@url [0]{\catcode `\\12\catcode `\$12\catcode `\&12\catcode `\#12\catcode `\^12\catcode `\_12\catcode `\%12\relax}%
\providecommand \@@startlink[1]{}%
\providecommand \@@endlink[0]{}%
\providecommand \url  [0]{\begingroup\@sanitize@url \@url }%
\providecommand \@url [1]{\endgroup\@href {#1}{\urlprefix }}%
\providecommand \urlprefix  [0]{URL }%
\providecommand \Eprint [0]{\href }%
\providecommand \doibase [0]{http://dx.doi.org/}%
\providecommand \selectlanguage [0]{\@gobble}%
\providecommand \bibinfo  [0]{\@secondoftwo}%
\providecommand \bibfield  [0]{\@secondoftwo}%
\providecommand \translation [1]{[#1]}%
\providecommand \BibitemOpen [0]{}%
\providecommand \bibitemStop [0]{}%
\providecommand \bibitemNoStop [0]{.\EOS\space}%
\providecommand \EOS [0]{\spacefactor3000\relax}%
\providecommand \BibitemShut  [1]{\csname bibitem#1\endcsname}%
\let\auto@bib@innerbib\@empty
\bibitem [{\citenamefont {Chen}\ \emph {et~al.}(2013{\natexlab{a}})\citenamefont {Chen}, \citenamefont {Gu}, \citenamefont {Liu},\ and\ \citenamefont {Wen}}]{wenspt}%
  \BibitemOpen
  \bibfield  {author} {\bibinfo {author} {\bibfnamefont {X.}~\bibnamefont {Chen}}, \bibinfo {author} {\bibfnamefont {Z.-C.}\ \bibnamefont {Gu}}, \bibinfo {author} {\bibfnamefont {Z.-X.}\ \bibnamefont {Liu}}, \ and\ \bibinfo {author} {\bibfnamefont {X.-G.}\ \bibnamefont {Wen}},\ }\href {\doibase 10.1103/PhysRevB.87.155114} {\bibfield  {journal} {\bibinfo  {journal} {Phys. Rev. B}\ }\textbf {\bibinfo {volume} {87}},\ \bibinfo {pages} {155114} (\bibinfo {year} {2013}{\natexlab{a}})}\BibitemShut {NoStop}%
\bibitem [{\citenamefont {Chen}\ \emph {et~al.}(2012)\citenamefont {Chen}, \citenamefont {Gu}, \citenamefont {Liu},\ and\ \citenamefont {Wen}}]{wenspt2}%
  \BibitemOpen
  \bibfield  {author} {\bibinfo {author} {\bibfnamefont {X.}~\bibnamefont {Chen}}, \bibinfo {author} {\bibfnamefont {Z.-C.}\ \bibnamefont {Gu}}, \bibinfo {author} {\bibfnamefont {Z.-X.}\ \bibnamefont {Liu}}, \ and\ \bibinfo {author} {\bibfnamefont {X.-G.}\ \bibnamefont {Wen}},\ }\href@noop {} {\bibfield  {journal} {\bibinfo  {journal} {Science}\ }\textbf {\bibinfo {volume} {338}},\ \bibinfo {pages} {1604} (\bibinfo {year} {2012})}\BibitemShut {NoStop}%
\bibitem [{\citenamefont {Cardy}(1989)}]{cardybCFT1}%
  \BibitemOpen
  \bibfield  {author} {\bibinfo {author} {\bibfnamefont {J.~L.}\ \bibnamefont {Cardy}},\ }\href {\doibase https://doi.org/10.1016/0550-3213(89)90521-X} {\bibfield  {journal} {\bibinfo  {journal} {Nuclear Physics B}\ }\textbf {\bibinfo {volume} {324}},\ \bibinfo {pages} {581} (\bibinfo {year} {1989})}\BibitemShut {NoStop}%
\bibitem [{\citenamefont {Cardy}\ and\ \citenamefont {Lewellen}(1991)}]{cardybCFT2}%
  \BibitemOpen
  \bibfield  {author} {\bibinfo {author} {\bibfnamefont {J.~L.}\ \bibnamefont {Cardy}}\ and\ \bibinfo {author} {\bibfnamefont {D.~C.}\ \bibnamefont {Lewellen}},\ }\href {\doibase https://doi.org/10.1016/0370-2693(91)90828-E} {\bibfield  {journal} {\bibinfo  {journal} {Physics Letters B}\ }\textbf {\bibinfo {volume} {259}},\ \bibinfo {pages} {274} (\bibinfo {year} {1991})}\BibitemShut {NoStop}%
\bibitem [{\citenamefont {Cardy}(1984)}]{cardyboundary}%
  \BibitemOpen
  \bibfield  {author} {\bibinfo {author} {\bibfnamefont {J.~L.}\ \bibnamefont {Cardy}},\ }\href {\doibase https://doi.org/10.1016/0550-3213(84)90241-4} {\bibfield  {journal} {\bibinfo  {journal} {Nuclear Physics B}\ }\textbf {\bibinfo {volume} {240}},\ \bibinfo {pages} {514 } (\bibinfo {year} {1984})}\BibitemShut {NoStop}%
\bibitem [{\citenamefont {Lewellen}(1992)}]{LewellenBCCO}%
  \BibitemOpen
  \bibfield  {author} {\bibinfo {author} {\bibfnamefont {D.~C.}\ \bibnamefont {Lewellen}},\ }\href {\doibase https://doi.org/10.1016/0550-3213(92)90370-Q} {\bibfield  {journal} {\bibinfo  {journal} {Nuclear Physics B}\ }\textbf {\bibinfo {volume} {372}},\ \bibinfo {pages} {654} (\bibinfo {year} {1992})}\BibitemShut {NoStop}%
\bibitem [{\citenamefont {Metlitski}(2022)}]{maxboundary}%
  \BibitemOpen
  \bibfield  {author} {\bibinfo {author} {\bibfnamefont {M.~A.}\ \bibnamefont {Metlitski}},\ }\href {\doibase 10.21468/SciPostPhys.12.4.131} {\bibfield  {journal} {\bibinfo  {journal} {SciPost Phys.}\ }\textbf {\bibinfo {volume} {12}},\ \bibinfo {pages} {131} (\bibinfo {year} {2022})}\BibitemShut {NoStop}%
\bibitem [{\citenamefont {Padayasi}\ \emph {et~al.}(2022)\citenamefont {Padayasi}, \citenamefont {Krishnan}, \citenamefont {Metlitski}, \citenamefont {Gruzberg},\ and\ \citenamefont {Meineri}}]{maxboundary2}%
  \BibitemOpen
  \bibfield  {author} {\bibinfo {author} {\bibfnamefont {J.}~\bibnamefont {Padayasi}}, \bibinfo {author} {\bibfnamefont {A.}~\bibnamefont {Krishnan}}, \bibinfo {author} {\bibfnamefont {M.~A.}\ \bibnamefont {Metlitski}}, \bibinfo {author} {\bibfnamefont {I.~A.}\ \bibnamefont {Gruzberg}}, \ and\ \bibinfo {author} {\bibfnamefont {M.}~\bibnamefont {Meineri}},\ }\href {\doibase 10.21468/SciPostPhys.12.6.190} {\bibfield  {journal} {\bibinfo  {journal} {SciPost Phys.}\ }\textbf {\bibinfo {volume} {12}},\ \bibinfo {pages} {190} (\bibinfo {year} {2022})}\BibitemShut {NoStop}%
\bibitem [{\citenamefont {Parisen~Toldin}(2021)}]{toldin1}%
  \BibitemOpen
  \bibfield  {author} {\bibinfo {author} {\bibfnamefont {F.}~\bibnamefont {Parisen~Toldin}},\ }\href {\doibase 10.1103/PhysRevLett.126.135701} {\bibfield  {journal} {\bibinfo  {journal} {Phys. Rev. Lett.}\ }\textbf {\bibinfo {volume} {126}},\ \bibinfo {pages} {135701} (\bibinfo {year} {2021})}\BibitemShut {NoStop}%
\bibitem [{\citenamefont {Parisen~Toldin}\ and\ \citenamefont {Metlitski}(2022)}]{toldin2}%
  \BibitemOpen
  \bibfield  {author} {\bibinfo {author} {\bibfnamefont {F.}~\bibnamefont {Parisen~Toldin}}\ and\ \bibinfo {author} {\bibfnamefont {M.~A.}\ \bibnamefont {Metlitski}},\ }\href {\doibase 10.1103/PhysRevLett.128.215701} {\bibfield  {journal} {\bibinfo  {journal} {Phys. Rev. Lett.}\ }\textbf {\bibinfo {volume} {128}},\ \bibinfo {pages} {215701} (\bibinfo {year} {2022})}\BibitemShut {NoStop}%
\bibitem [{\citenamefont {Hu}\ \emph {et~al.}(2021)\citenamefont {Hu}, \citenamefont {Deng},\ and\ \citenamefont {Lv}}]{lvboundary}%
  \BibitemOpen
  \bibfield  {author} {\bibinfo {author} {\bibfnamefont {M.}~\bibnamefont {Hu}}, \bibinfo {author} {\bibfnamefont {Y.}~\bibnamefont {Deng}}, \ and\ \bibinfo {author} {\bibfnamefont {J.-P.}\ \bibnamefont {Lv}},\ }\href {\doibase 10.1103/PhysRevLett.127.120603} {\bibfield  {journal} {\bibinfo  {journal} {Phys. Rev. Lett.}\ }\textbf {\bibinfo {volume} {127}},\ \bibinfo {pages} {120603} (\bibinfo {year} {2021})}\BibitemShut {NoStop}%
\bibitem [{\citenamefont {Zhang}\ and\ \citenamefont {Wang}(2017)}]{zhang1}%
  \BibitemOpen
  \bibfield  {author} {\bibinfo {author} {\bibfnamefont {L.}~\bibnamefont {Zhang}}\ and\ \bibinfo {author} {\bibfnamefont {F.}~\bibnamefont {Wang}},\ }\href {\doibase 10.1103/PhysRevLett.118.087201} {\bibfield  {journal} {\bibinfo  {journal} {Phys. Rev. Lett.}\ }\textbf {\bibinfo {volume} {118}},\ \bibinfo {pages} {087201} (\bibinfo {year} {2017})}\BibitemShut {NoStop}%
\bibitem [{\citenamefont {Ding}\ \emph {et~al.}(2018)\citenamefont {Ding}, \citenamefont {Zhang},\ and\ \citenamefont {Guo}}]{zhang2}%
  \BibitemOpen
  \bibfield  {author} {\bibinfo {author} {\bibfnamefont {C.}~\bibnamefont {Ding}}, \bibinfo {author} {\bibfnamefont {L.}~\bibnamefont {Zhang}}, \ and\ \bibinfo {author} {\bibfnamefont {W.}~\bibnamefont {Guo}},\ }\href {\doibase 10.1103/PhysRevLett.120.235701} {\bibfield  {journal} {\bibinfo  {journal} {Phys. Rev. Lett.}\ }\textbf {\bibinfo {volume} {120}},\ \bibinfo {pages} {235701} (\bibinfo {year} {2018})}\BibitemShut {NoStop}%
\bibitem [{\citenamefont {Weber}\ \emph {et~al.}(2018)\citenamefont {Weber}, \citenamefont {Parisen~Toldin},\ and\ \citenamefont {Wessel}}]{stefan1}%
  \BibitemOpen
  \bibfield  {author} {\bibinfo {author} {\bibfnamefont {L.}~\bibnamefont {Weber}}, \bibinfo {author} {\bibfnamefont {F.}~\bibnamefont {Parisen~Toldin}}, \ and\ \bibinfo {author} {\bibfnamefont {S.}~\bibnamefont {Wessel}},\ }\href {\doibase 10.1103/PhysRevB.98.140403} {\bibfield  {journal} {\bibinfo  {journal} {Phys. Rev. B}\ }\textbf {\bibinfo {volume} {98}},\ \bibinfo {pages} {140403} (\bibinfo {year} {2018})}\BibitemShut {NoStop}%
\bibitem [{\citenamefont {Weber}\ and\ \citenamefont {Wessel}(2019)}]{stefan2}%
  \BibitemOpen
  \bibfield  {author} {\bibinfo {author} {\bibfnamefont {L.}~\bibnamefont {Weber}}\ and\ \bibinfo {author} {\bibfnamefont {S.}~\bibnamefont {Wessel}},\ }\href {\doibase 10.1103/PhysRevB.100.054437} {\bibfield  {journal} {\bibinfo  {journal} {Phys. Rev. B}\ }\textbf {\bibinfo {volume} {100}},\ \bibinfo {pages} {054437} (\bibinfo {year} {2019})}\BibitemShut {NoStop}%
\bibitem [{\citenamefont {{Grover}}\ and\ \citenamefont {{Vishwanath}}(2012)}]{groveredge}%
  \BibitemOpen
  \bibfield  {author} {\bibinfo {author} {\bibfnamefont {T.}~\bibnamefont {{Grover}}}\ and\ \bibinfo {author} {\bibfnamefont {A.}~\bibnamefont {{Vishwanath}}},\ }\href@noop {} {\bibfield  {journal} {\bibinfo  {journal} {arXiv e-prints}\ ,\ \bibinfo {eid} {arXiv:1206.1332}} (\bibinfo {year} {2012})},\ \Eprint {http://arxiv.org/abs/1206.1332} {arXiv:1206.1332 [cond-mat.str-el]} \BibitemShut {NoStop}%
\bibitem [{\citenamefont {Xu}\ \emph {et~al.}(2020)\citenamefont {Xu}, \citenamefont {Wu}, \citenamefont {Jian},\ and\ \citenamefont {Xu}}]{edgexu1}%
  \BibitemOpen
  \bibfield  {author} {\bibinfo {author} {\bibfnamefont {Y.}~\bibnamefont {Xu}}, \bibinfo {author} {\bibfnamefont {X.-C.}\ \bibnamefont {Wu}}, \bibinfo {author} {\bibfnamefont {C.-M.}\ \bibnamefont {Jian}}, \ and\ \bibinfo {author} {\bibfnamefont {C.}~\bibnamefont {Xu}},\ }\href {\doibase 10.1103/PhysRevB.101.184419} {\bibfield  {journal} {\bibinfo  {journal} {Phys. Rev. B}\ }\textbf {\bibinfo {volume} {101}},\ \bibinfo {pages} {184419} (\bibinfo {year} {2020})}\BibitemShut {NoStop}%
\bibitem [{\citenamefont {Jian}\ \emph {et~al.}(2021)\citenamefont {Jian}, \citenamefont {Xu}, \citenamefont {Wu},\ and\ \citenamefont {Xu}}]{edgexu2}%
  \BibitemOpen
  \bibfield  {author} {\bibinfo {author} {\bibfnamefont {C.-M.}\ \bibnamefont {Jian}}, \bibinfo {author} {\bibfnamefont {Y.}~\bibnamefont {Xu}}, \bibinfo {author} {\bibfnamefont {X.-C.}\ \bibnamefont {Wu}}, \ and\ \bibinfo {author} {\bibfnamefont {C.}~\bibnamefont {Xu}},\ }\href {\doibase 10.21468/SciPostPhys.10.2.033} {\bibfield  {journal} {\bibinfo  {journal} {SciPost Phys.}\ }\textbf {\bibinfo {volume} {10}},\ \bibinfo {pages} {33} (\bibinfo {year} {2021})}\BibitemShut {NoStop}%
\bibitem [{\citenamefont {Zhang}\ \emph {et~al.}(2023)\citenamefont {Zhang}, \citenamefont {Zhu},\ and\ \citenamefont {Vishwanath}}]{ZhangFQSHExtralog}%
  \BibitemOpen
  \bibfield  {author} {\bibinfo {author} {\bibfnamefont {Y.-H.}\ \bibnamefont {Zhang}}, \bibinfo {author} {\bibfnamefont {Z.}~\bibnamefont {Zhu}}, \ and\ \bibinfo {author} {\bibfnamefont {A.}~\bibnamefont {Vishwanath}},\ }\href {\doibase 10.1103/PhysRevX.13.031023} {\bibfield  {journal} {\bibinfo  {journal} {Phys. Rev. X}\ }\textbf {\bibinfo {volume} {13}},\ \bibinfo {pages} {031023} (\bibinfo {year} {2023})}\BibitemShut {NoStop}%
\bibitem [{\citenamefont {Thorngren}\ and\ \citenamefont {Wang}(2021)}]{ThorngrenWanganomalousbdry}%
  \BibitemOpen
  \bibfield  {author} {\bibinfo {author} {\bibfnamefont {R.}~\bibnamefont {Thorngren}}\ and\ \bibinfo {author} {\bibfnamefont {Y.}~\bibnamefont {Wang}},\ }\href {\doibase 10.1007/JHEP09(2021)017} {\bibfield  {journal} {\bibinfo  {journal} {Journal of High Energy Physics}\ }\textbf {\bibinfo {volume} {2021}},\ \bibinfo {pages} {17} (\bibinfo {year} {2021})}\BibitemShut {NoStop}%
\bibitem [{\citenamefont {Han}\ \emph {et~al.}(2017)\citenamefont {Han}, \citenamefont {Tiwari}, \citenamefont {Hsieh},\ and\ \citenamefont {Ryu}}]{Ryuanomalousbdry}%
  \BibitemOpen
  \bibfield  {author} {\bibinfo {author} {\bibfnamefont {B.}~\bibnamefont {Han}}, \bibinfo {author} {\bibfnamefont {A.}~\bibnamefont {Tiwari}}, \bibinfo {author} {\bibfnamefont {C.-T.}\ \bibnamefont {Hsieh}}, \ and\ \bibinfo {author} {\bibfnamefont {S.}~\bibnamefont {Ryu}},\ }\href {\doibase 10.1103/PhysRevB.96.125105} {\bibfield  {journal} {\bibinfo  {journal} {Phys. Rev. B}\ }\textbf {\bibinfo {volume} {96}},\ \bibinfo {pages} {125105} (\bibinfo {year} {2017})}\BibitemShut {NoStop}%
\bibitem [{\citenamefont {Senthil}\ \emph {et~al.}(2004{\natexlab{a}})\citenamefont {Senthil}, \citenamefont {Vishwanath}, \citenamefont {Balents}, \citenamefont {Sachdev},\ and\ \citenamefont {Fisher}}]{deconfine1}%
  \BibitemOpen
  \bibfield  {author} {\bibinfo {author} {\bibfnamefont {T.}~\bibnamefont {Senthil}}, \bibinfo {author} {\bibfnamefont {A.}~\bibnamefont {Vishwanath}}, \bibinfo {author} {\bibfnamefont {L.}~\bibnamefont {Balents}}, \bibinfo {author} {\bibfnamefont {S.}~\bibnamefont {Sachdev}}, \ and\ \bibinfo {author} {\bibfnamefont {M.~P.~A.}\ \bibnamefont {Fisher}},\ }\href@noop {} {\bibfield  {journal} {\bibinfo  {journal} {Science}\ }\textbf {\bibinfo {volume} {303}},\ \bibinfo {pages} {1490} (\bibinfo {year} {2004}{\natexlab{a}})}\BibitemShut {NoStop}%
\bibitem [{\citenamefont {Senthil}\ \emph {et~al.}(2004{\natexlab{b}})\citenamefont {Senthil}, \citenamefont {Balents}, \citenamefont {Sachdev}, \citenamefont {Vishwanath},\ and\ \citenamefont {Fisher}}]{deconfine2}%
  \BibitemOpen
  \bibfield  {author} {\bibinfo {author} {\bibfnamefont {T.}~\bibnamefont {Senthil}}, \bibinfo {author} {\bibfnamefont {L.}~\bibnamefont {Balents}}, \bibinfo {author} {\bibfnamefont {S.}~\bibnamefont {Sachdev}}, \bibinfo {author} {\bibfnamefont {A.}~\bibnamefont {Vishwanath}}, \ and\ \bibinfo {author} {\bibfnamefont {M.~P.~A.}\ \bibnamefont {Fisher}},\ }\href {\doibase 10.1103/PhysRevB.70.144407} {\bibfield  {journal} {\bibinfo  {journal} {Phys. Rev. B}\ }\textbf {\bibinfo {volume} {70}},\ \bibinfo {pages} {144407} (\bibinfo {year} {2004}{\natexlab{b}})}\BibitemShut {NoStop}%
\bibitem [{\citenamefont {Vishwanath}\ and\ \citenamefont {Senthil}(2013)}]{senthilashvin}%
  \BibitemOpen
  \bibfield  {author} {\bibinfo {author} {\bibfnamefont {A.}~\bibnamefont {Vishwanath}}\ and\ \bibinfo {author} {\bibfnamefont {T.}~\bibnamefont {Senthil}},\ }\href {\doibase 10.1103/PhysRevX.3.011016} {\bibfield  {journal} {\bibinfo  {journal} {Phys. Rev. X}\ }\textbf {\bibinfo {volume} {3}},\ \bibinfo {pages} {011016} (\bibinfo {year} {2013})}\BibitemShut {NoStop}%
\bibitem [{\citenamefont {Wang}\ \emph {et~al.}(2017)\citenamefont {Wang}, \citenamefont {Nahum}, \citenamefont {Metlitski}, \citenamefont {Xu},\ and\ \citenamefont {Senthil}}]{SO5}%
  \BibitemOpen
  \bibfield  {author} {\bibinfo {author} {\bibfnamefont {C.}~\bibnamefont {Wang}}, \bibinfo {author} {\bibfnamefont {A.}~\bibnamefont {Nahum}}, \bibinfo {author} {\bibfnamefont {M.~A.}\ \bibnamefont {Metlitski}}, \bibinfo {author} {\bibfnamefont {C.}~\bibnamefont {Xu}}, \ and\ \bibinfo {author} {\bibfnamefont {T.}~\bibnamefont {Senthil}},\ }\href {\doibase 10.1103/PhysRevX.7.031051} {\bibfield  {journal} {\bibinfo  {journal} {Phys. Rev. X}\ }\textbf {\bibinfo {volume} {7}},\ \bibinfo {pages} {031051} (\bibinfo {year} {2017})}\BibitemShut {NoStop}%
\bibitem [{\citenamefont {Keselman}\ and\ \citenamefont {Berg}(2015)}]{gspt1}%
  \BibitemOpen
  \bibfield  {author} {\bibinfo {author} {\bibfnamefont {A.}~\bibnamefont {Keselman}}\ and\ \bibinfo {author} {\bibfnamefont {E.}~\bibnamefont {Berg}},\ }\href {\doibase 10.1103/PhysRevB.91.235309} {\bibfield  {journal} {\bibinfo  {journal} {Phys. Rev. B}\ }\textbf {\bibinfo {volume} {91}},\ \bibinfo {pages} {235309} (\bibinfo {year} {2015})}\BibitemShut {NoStop}%
\bibitem [{\citenamefont {Scaffidi}\ \emph {et~al.}(2017)\citenamefont {Scaffidi}, \citenamefont {Parker},\ and\ \citenamefont {Vasseur}}]{gspt2}%
  \BibitemOpen
  \bibfield  {author} {\bibinfo {author} {\bibfnamefont {T.}~\bibnamefont {Scaffidi}}, \bibinfo {author} {\bibfnamefont {D.~E.}\ \bibnamefont {Parker}}, \ and\ \bibinfo {author} {\bibfnamefont {R.}~\bibnamefont {Vasseur}},\ }\href {\doibase 10.1103/PhysRevX.7.041048} {\bibfield  {journal} {\bibinfo  {journal} {Phys. Rev. X}\ }\textbf {\bibinfo {volume} {7}},\ \bibinfo {pages} {041048} (\bibinfo {year} {2017})}\BibitemShut {NoStop}%
\bibitem [{\citenamefont {Verresen}\ \emph {et~al.}(2018)\citenamefont {Verresen}, \citenamefont {Jones},\ and\ \citenamefont {Pollmann}}]{gspt3}%
  \BibitemOpen
  \bibfield  {author} {\bibinfo {author} {\bibfnamefont {R.}~\bibnamefont {Verresen}}, \bibinfo {author} {\bibfnamefont {N.~G.}\ \bibnamefont {Jones}}, \ and\ \bibinfo {author} {\bibfnamefont {F.}~\bibnamefont {Pollmann}},\ }\href {\doibase 10.1103/PhysRevLett.120.057001} {\bibfield  {journal} {\bibinfo  {journal} {Phys. Rev. Lett.}\ }\textbf {\bibinfo {volume} {120}},\ \bibinfo {pages} {057001} (\bibinfo {year} {2018})}\BibitemShut {NoStop}%
\bibitem [{\citenamefont {Verresen}\ \emph {et~al.}(2021)\citenamefont {Verresen}, \citenamefont {Thorngren}, \citenamefont {Jones},\ and\ \citenamefont {Pollmann}}]{gspt4}%
  \BibitemOpen
  \bibfield  {author} {\bibinfo {author} {\bibfnamefont {R.}~\bibnamefont {Verresen}}, \bibinfo {author} {\bibfnamefont {R.}~\bibnamefont {Thorngren}}, \bibinfo {author} {\bibfnamefont {N.~G.}\ \bibnamefont {Jones}}, \ and\ \bibinfo {author} {\bibfnamefont {F.}~\bibnamefont {Pollmann}},\ }\href {\doibase 10.1103/PhysRevX.11.041059} {\bibfield  {journal} {\bibinfo  {journal} {Phys. Rev. X}\ }\textbf {\bibinfo {volume} {11}},\ \bibinfo {pages} {041059} (\bibinfo {year} {2021})}\BibitemShut {NoStop}%
\bibitem [{\citenamefont {Thorngren}\ \emph {et~al.}(2021)\citenamefont {Thorngren}, \citenamefont {Vishwanath},\ and\ \citenamefont {Verresen}}]{igspt1}%
  \BibitemOpen
  \bibfield  {author} {\bibinfo {author} {\bibfnamefont {R.}~\bibnamefont {Thorngren}}, \bibinfo {author} {\bibfnamefont {A.}~\bibnamefont {Vishwanath}}, \ and\ \bibinfo {author} {\bibfnamefont {R.}~\bibnamefont {Verresen}},\ }\href {\doibase 10.1103/PhysRevB.104.075132} {\bibfield  {journal} {\bibinfo  {journal} {Phys. Rev. B}\ }\textbf {\bibinfo {volume} {104}},\ \bibinfo {pages} {075132} (\bibinfo {year} {2021})}\BibitemShut {NoStop}%
\bibitem [{\citenamefont {Wen}\ and\ \citenamefont {Potter}(2023)}]{igspt2}%
  \BibitemOpen
  \bibfield  {author} {\bibinfo {author} {\bibfnamefont {R.}~\bibnamefont {Wen}}\ and\ \bibinfo {author} {\bibfnamefont {A.~C.}\ \bibnamefont {Potter}},\ }\href {\doibase 10.1103/PhysRevB.107.245127} {\bibfield  {journal} {\bibinfo  {journal} {Phys. Rev. B}\ }\textbf {\bibinfo {volume} {107}},\ \bibinfo {pages} {245127} (\bibinfo {year} {2023})}\BibitemShut {NoStop}%
\bibitem [{\citenamefont {Grover}\ and\ \citenamefont {Senthil}(2008)}]{groversenthil}%
  \BibitemOpen
  \bibfield  {author} {\bibinfo {author} {\bibfnamefont {T.}~\bibnamefont {Grover}}\ and\ \bibinfo {author} {\bibfnamefont {T.}~\bibnamefont {Senthil}},\ }\href {\doibase 10.1103/PhysRevLett.100.156804} {\bibfield  {journal} {\bibinfo  {journal} {Phys. Rev. Lett.}\ }\textbf {\bibinfo {volume} {100}},\ \bibinfo {pages} {156804} (\bibinfo {year} {2008})}\BibitemShut {NoStop}%
\bibitem [{\citenamefont {Liu}\ \emph {et~al.}(2019)\citenamefont {Liu}, \citenamefont {Wang}, \citenamefont {Sato}, \citenamefont {Hohenadler}, \citenamefont {Wang}, \citenamefont {Guo},\ and\ \citenamefont {Assaad}}]{qshscfakher}%
  \BibitemOpen
  \bibfield  {author} {\bibinfo {author} {\bibfnamefont {Y.}~\bibnamefont {Liu}}, \bibinfo {author} {\bibfnamefont {Z.}~\bibnamefont {Wang}}, \bibinfo {author} {\bibfnamefont {T.}~\bibnamefont {Sato}}, \bibinfo {author} {\bibfnamefont {M.}~\bibnamefont {Hohenadler}}, \bibinfo {author} {\bibfnamefont {C.}~\bibnamefont {Wang}}, \bibinfo {author} {\bibfnamefont {W.}~\bibnamefont {Guo}}, \ and\ \bibinfo {author} {\bibfnamefont {F.~F.}\ \bibnamefont {Assaad}},\ }\href {\doibase 10.1038/s41467-019-10372-0} {\bibfield  {journal} {\bibinfo  {journal} {Nature Communications}\ }\textbf {\bibinfo {volume} {10}} (\bibinfo {year} {2019}),\ 10.1038/s41467-019-10372-0}\BibitemShut {NoStop}%
\bibitem [{\citenamefont {Liu}\ \emph {et~al.}(2023)\citenamefont {Liu}, \citenamefont {Jiang}, \citenamefont {Chen}, \citenamefont {Rong}, \citenamefont {Cheng}, \citenamefont {Sun}, \citenamefont {Meng},\ and\ \citenamefont {Assaad}}]{qshscfakher2}%
  \BibitemOpen
  \bibfield  {author} {\bibinfo {author} {\bibfnamefont {Z.~H.}\ \bibnamefont {Liu}}, \bibinfo {author} {\bibfnamefont {W.}~\bibnamefont {Jiang}}, \bibinfo {author} {\bibfnamefont {B.-B.}\ \bibnamefont {Chen}}, \bibinfo {author} {\bibfnamefont {J.}~\bibnamefont {Rong}}, \bibinfo {author} {\bibfnamefont {M.}~\bibnamefont {Cheng}}, \bibinfo {author} {\bibfnamefont {K.}~\bibnamefont {Sun}}, \bibinfo {author} {\bibfnamefont {Z.~Y.}\ \bibnamefont {Meng}}, \ and\ \bibinfo {author} {\bibfnamefont {F.~F.}\ \bibnamefont {Assaad}},\ }\href {\doibase 10.1103/PhysRevLett.130.266501} {\bibfield  {journal} {\bibinfo  {journal} {Phys. Rev. Lett.}\ }\textbf {\bibinfo {volume} {130}},\ \bibinfo {pages} {266501} (\bibinfo {year} {2023})}\BibitemShut {NoStop}%
\bibitem [{\citenamefont {Khalaf}\ \emph {et~al.}(2021)\citenamefont {Khalaf}, \citenamefont {Chatterjee}, \citenamefont {Bultinck}, \citenamefont {Zaletel},\ and\ \citenamefont {Vishwanath}}]{skyrmionsc}%
  \BibitemOpen
  \bibfield  {author} {\bibinfo {author} {\bibfnamefont {E.}~\bibnamefont {Khalaf}}, \bibinfo {author} {\bibfnamefont {S.}~\bibnamefont {Chatterjee}}, \bibinfo {author} {\bibfnamefont {N.}~\bibnamefont {Bultinck}}, \bibinfo {author} {\bibfnamefont {M.~P.}\ \bibnamefont {Zaletel}}, \ and\ \bibinfo {author} {\bibfnamefont {A.}~\bibnamefont {Vishwanath}},\ }\href {\doibase 10.1126/sciadv.abf5299} {\bibfield  {journal} {\bibinfo  {journal} {Science Advances}\ }\textbf {\bibinfo {volume} {7}} (\bibinfo {year} {2021}),\ 10.1126/sciadv.abf5299}\BibitemShut {NoStop}%
\bibitem [{\citenamefont {Assaad}(2024)}]{fakhertalk}%
  \BibitemOpen
  \bibfield  {author} {\bibinfo {author} {\bibfnamefont {F.}~\bibnamefont {Assaad}},\ }\href {https://online.kitp.ucsb.edu/online/gapless24/assaad/} {\enquote {\bibinfo {title} {Fermion models of deconfined quantum criticality},}\ } (\bibinfo {year} {2024}),\ \bibinfo {note} {kITP Program: Correlated Gapless Quantum Matter}\BibitemShut {NoStop}%
\bibitem [{\citenamefont {Chen}\ \emph {et~al.}(2013{\natexlab{b}})\citenamefont {Chen}, \citenamefont {Huang}, \citenamefont {Deng}, \citenamefont {Kuklov}, \citenamefont {Prokof'ev},\ and\ \citenamefont {Svistunov}}]{1storder}%
  \BibitemOpen
  \bibfield  {author} {\bibinfo {author} {\bibfnamefont {K.}~\bibnamefont {Chen}}, \bibinfo {author} {\bibfnamefont {Y.}~\bibnamefont {Huang}}, \bibinfo {author} {\bibfnamefont {Y.}~\bibnamefont {Deng}}, \bibinfo {author} {\bibfnamefont {A.~B.}\ \bibnamefont {Kuklov}}, \bibinfo {author} {\bibfnamefont {N.~V.}\ \bibnamefont {Prokof'ev}}, \ and\ \bibinfo {author} {\bibfnamefont {B.~V.}\ \bibnamefont {Svistunov}},\ }\href {\doibase 10.1103/PhysRevLett.110.185701} {\bibfield  {journal} {\bibinfo  {journal} {Phys. Rev. Lett.}\ }\textbf {\bibinfo {volume} {110}},\ \bibinfo {pages} {185701} (\bibinfo {year} {2013}{\natexlab{b}})}\BibitemShut {NoStop}%
\bibitem [{\citenamefont {Nahum}\ \emph {et~al.}(2015{\natexlab{a}})\citenamefont {Nahum}, \citenamefont {Serna}, \citenamefont {Chalker}, \citenamefont {Ortu\~no},\ and\ \citenamefont {Somoza}}]{loopmodel1}%
  \BibitemOpen
  \bibfield  {author} {\bibinfo {author} {\bibfnamefont {A.}~\bibnamefont {Nahum}}, \bibinfo {author} {\bibfnamefont {P.}~\bibnamefont {Serna}}, \bibinfo {author} {\bibfnamefont {J.~T.}\ \bibnamefont {Chalker}}, \bibinfo {author} {\bibfnamefont {M.}~\bibnamefont {Ortu\~no}}, \ and\ \bibinfo {author} {\bibfnamefont {A.~M.}\ \bibnamefont {Somoza}},\ }\href {\doibase 10.1103/PhysRevLett.115.267203} {\bibfield  {journal} {\bibinfo  {journal} {Phys. Rev. Lett.}\ }\textbf {\bibinfo {volume} {115}},\ \bibinfo {pages} {267203} (\bibinfo {year} {2015}{\natexlab{a}})}\BibitemShut {NoStop}%
\bibitem [{\citenamefont {Nahum}\ \emph {et~al.}(2015{\natexlab{b}})\citenamefont {Nahum}, \citenamefont {Chalker}, \citenamefont {Serna}, \citenamefont {Ortu\~no},\ and\ \citenamefont {Somoza}}]{loopmodel2}%
  \BibitemOpen
  \bibfield  {author} {\bibinfo {author} {\bibfnamefont {A.}~\bibnamefont {Nahum}}, \bibinfo {author} {\bibfnamefont {J.~T.}\ \bibnamefont {Chalker}}, \bibinfo {author} {\bibfnamefont {P.}~\bibnamefont {Serna}}, \bibinfo {author} {\bibfnamefont {M.}~\bibnamefont {Ortu\~no}}, \ and\ \bibinfo {author} {\bibfnamefont {A.~M.}\ \bibnamefont {Somoza}},\ }\href {\doibase 10.1103/PhysRevX.5.041048} {\bibfield  {journal} {\bibinfo  {journal} {Phys. Rev. X}\ }\textbf {\bibinfo {volume} {5}},\ \bibinfo {pages} {041048} (\bibinfo {year} {2015}{\natexlab{b}})}\BibitemShut {NoStop}%
\bibitem [{\citenamefont {Poland}\ \emph {et~al.}(2019)\citenamefont {Poland}, \citenamefont {Rychkov},\ and\ \citenamefont {Vichi}}]{bootstrap}%
  \BibitemOpen
  \bibfield  {author} {\bibinfo {author} {\bibfnamefont {D.}~\bibnamefont {Poland}}, \bibinfo {author} {\bibfnamefont {S.}~\bibnamefont {Rychkov}}, \ and\ \bibinfo {author} {\bibfnamefont {A.}~\bibnamefont {Vichi}},\ }\href {\doibase 10.1103/RevModPhys.91.015002} {\bibfield  {journal} {\bibinfo  {journal} {Rev. Mod. Phys.}\ }\textbf {\bibinfo {volume} {91}},\ \bibinfo {pages} {015002} (\bibinfo {year} {2019})}\BibitemShut {NoStop}%
\bibitem [{\citenamefont {Nakayama}\ and\ \citenamefont {Ohtsuki}(2016)}]{bootstrap1}%
  \BibitemOpen
  \bibfield  {author} {\bibinfo {author} {\bibfnamefont {Y.}~\bibnamefont {Nakayama}}\ and\ \bibinfo {author} {\bibfnamefont {T.}~\bibnamefont {Ohtsuki}},\ }\href {\doibase 10.1103/PhysRevLett.117.131601} {\bibfield  {journal} {\bibinfo  {journal} {Phys. Rev. Lett.}\ }\textbf {\bibinfo {volume} {117}},\ \bibinfo {pages} {131601} (\bibinfo {year} {2016})}\BibitemShut {NoStop}%
\bibitem [{\citenamefont {Song}\ \emph {et~al.}(2024{\natexlab{a}})\citenamefont {Song}, \citenamefont {Zhao}, \citenamefont {Meng}, \citenamefont {Xu},\ and\ \citenamefont {Cheng}}]{song2024extracting}%
  \BibitemOpen
  \bibfield  {author} {\bibinfo {author} {\bibfnamefont {M.}~\bibnamefont {Song}}, \bibinfo {author} {\bibfnamefont {J.}~\bibnamefont {Zhao}}, \bibinfo {author} {\bibfnamefont {Z.~Y.}\ \bibnamefont {Meng}}, \bibinfo {author} {\bibfnamefont {C.}~\bibnamefont {Xu}}, \ and\ \bibinfo {author} {\bibfnamefont {M.}~\bibnamefont {Cheng}},\ }\href@noop {} {\enquote {\bibinfo {title} {Extracting subleading corrections in entanglement entropy at quantum phase transitions},}\ } (\bibinfo {year} {2024}{\natexlab{a}}),\ \Eprint {http://arxiv.org/abs/2312.13498} {arXiv:2312.13498 [cond-mat.str-el]} \BibitemShut {NoStop}%
\bibitem [{\citenamefont {Song}\ \emph {et~al.}(2024{\natexlab{b}})\citenamefont {Song}, \citenamefont {Zhao}, \citenamefont {Cheng}, \citenamefont {Xu}, \citenamefont {Scherer}, \citenamefont {Janssen},\ and\ \citenamefont {Meng}}]{song2024deconfined}%
  \BibitemOpen
  \bibfield  {author} {\bibinfo {author} {\bibfnamefont {M.}~\bibnamefont {Song}}, \bibinfo {author} {\bibfnamefont {J.}~\bibnamefont {Zhao}}, \bibinfo {author} {\bibfnamefont {M.}~\bibnamefont {Cheng}}, \bibinfo {author} {\bibfnamefont {C.}~\bibnamefont {Xu}}, \bibinfo {author} {\bibfnamefont {M.~M.}\ \bibnamefont {Scherer}}, \bibinfo {author} {\bibfnamefont {L.}~\bibnamefont {Janssen}}, \ and\ \bibinfo {author} {\bibfnamefont {Z.~Y.}\ \bibnamefont {Meng}},\ }\href@noop {} {\enquote {\bibinfo {title} {Deconfined quantum criticality lost},}\ } (\bibinfo {year} {2024}{\natexlab{b}}),\ \Eprint {http://arxiv.org/abs/2307.02547} {arXiv:2307.02547 [cond-mat.str-el]} \BibitemShut {NoStop}%
\bibitem [{\citenamefont {Zhou}\ \emph {et~al.}(2024)\citenamefont {Zhou}, \citenamefont {Hu}, \citenamefont {Zhu},\ and\ \citenamefont {He}}]{hefuzzy}%
  \BibitemOpen
  \bibfield  {author} {\bibinfo {author} {\bibfnamefont {Z.}~\bibnamefont {Zhou}}, \bibinfo {author} {\bibfnamefont {L.}~\bibnamefont {Hu}}, \bibinfo {author} {\bibfnamefont {W.}~\bibnamefont {Zhu}}, \ and\ \bibinfo {author} {\bibfnamefont {Y.-C.}\ \bibnamefont {He}},\ }\href@noop {} {\enquote {\bibinfo {title} {The $\mathrm{SO}(5)$ deconfined phase transition under the fuzzy sphere microscope: Approximate conformal symmetry, pseudo-criticality, and operator spectrum},}\ } (\bibinfo {year} {2024}),\ \Eprint {http://arxiv.org/abs/2306.16435} {arXiv:2306.16435 [cond-mat.str-el]} \BibitemShut {NoStop}%
\bibitem [{\citenamefont {Takahashi}\ \emph {et~al.}(2024)\citenamefont {Takahashi}, \citenamefont {Shao}, \citenamefont {Zhao}, \citenamefont {Guo},\ and\ \citenamefont {Sandvik}}]{sandvik1st}%
  \BibitemOpen
  \bibfield  {author} {\bibinfo {author} {\bibfnamefont {J.}~\bibnamefont {Takahashi}}, \bibinfo {author} {\bibfnamefont {H.}~\bibnamefont {Shao}}, \bibinfo {author} {\bibfnamefont {B.}~\bibnamefont {Zhao}}, \bibinfo {author} {\bibfnamefont {W.}~\bibnamefont {Guo}}, \ and\ \bibinfo {author} {\bibfnamefont {A.~W.}\ \bibnamefont {Sandvik}},\ }\href@noop {} {\enquote {\bibinfo {title} {So(5) multicriticality in two-dimensional quantum magnets},}\ } (\bibinfo {year} {2024}),\ \Eprint {http://arxiv.org/abs/2405.06607} {arXiv:2405.06607 [cond-mat.str-el]} \BibitemShut {NoStop}%
\bibitem [{\citenamefont {Ma}\ \emph {et~al.}(2022)\citenamefont {Ma}, \citenamefont {Zou},\ and\ \citenamefont {Wang}}]{mazouwang}%
  \BibitemOpen
  \bibfield  {author} {\bibinfo {author} {\bibfnamefont {R.}~\bibnamefont {Ma}}, \bibinfo {author} {\bibfnamefont {L.}~\bibnamefont {Zou}}, \ and\ \bibinfo {author} {\bibfnamefont {C.}~\bibnamefont {Wang}},\ }\href {\doibase 10.21468/SciPostPhys.12.6.196} {\bibfield  {journal} {\bibinfo  {journal} {SciPost Phys.}\ }\textbf {\bibinfo {volume} {12}},\ \bibinfo {pages} {196} (\bibinfo {year} {2022})}\BibitemShut {NoStop}%
\bibitem [{\citenamefont {Kane}\ and\ \citenamefont {Mele}(2005{\natexlab{a}})}]{kane2005a}%
  \BibitemOpen
  \bibfield  {author} {\bibinfo {author} {\bibfnamefont {C.~L.}\ \bibnamefont {Kane}}\ and\ \bibinfo {author} {\bibfnamefont {E.~J.}\ \bibnamefont {Mele}},\ }\href@noop {} {\bibfield  {journal} {\bibinfo  {journal} {Physical Review Letter}\ }\textbf {\bibinfo {volume} {95}},\ \bibinfo {pages} {226801} (\bibinfo {year} {2005}{\natexlab{a}})}\BibitemShut {NoStop}%
\bibitem [{\citenamefont {Kane}\ and\ \citenamefont {Mele}(2005{\natexlab{b}})}]{kane2005b}%
  \BibitemOpen
  \bibfield  {author} {\bibinfo {author} {\bibfnamefont {C.~L.}\ \bibnamefont {Kane}}\ and\ \bibinfo {author} {\bibfnamefont {E.~J.}\ \bibnamefont {Mele}},\ }\href@noop {} {\bibfield  {journal} {\bibinfo  {journal} {Physical Review Letter}\ }\textbf {\bibinfo {volume} {95}},\ \bibinfo {pages} {146802} (\bibinfo {year} {2005}{\natexlab{b}})}\BibitemShut {NoStop}%
\bibitem [{\citenamefont {Cardy}(1996)}]{cardybook}%
  \BibitemOpen
  \bibfield  {author} {\bibinfo {author} {\bibfnamefont {J.}~\bibnamefont {Cardy}},\ }\href@noop {} {\emph {\bibinfo {title} {Scaling and Renormalization in Statistical Physics}}}\ (\bibinfo  {publisher} {Cambridge Lecture Notes in Physics},\ \bibinfo {year} {1996})\BibitemShut {NoStop}%
\bibitem [{\citenamefont {Dietrich}\ and\ \citenamefont {Diehl}(1983)}]{boundary2}%
  \BibitemOpen
  \bibfield  {author} {\bibinfo {author} {\bibfnamefont {S.}~\bibnamefont {Dietrich}}\ and\ \bibinfo {author} {\bibfnamefont {H.~W.}\ \bibnamefont {Diehl}},\ }\href {\doibase 10.1007/BF01319217} {\bibfield  {journal} {\bibinfo  {journal} {Zeitschrift f{\"u}r Physik B Condensed Matter}\ }\textbf {\bibinfo {volume} {51}},\ \bibinfo {pages} {343} (\bibinfo {year} {1983})}\BibitemShut {NoStop}%
\bibitem [{\citenamefont {Diehl}\ and\ \citenamefont {Dietrich}(1981)}]{boundary3}%
  \BibitemOpen
  \bibfield  {author} {\bibinfo {author} {\bibfnamefont {H.~W.}\ \bibnamefont {Diehl}}\ and\ \bibinfo {author} {\bibfnamefont {S.}~\bibnamefont {Dietrich}},\ }\href {\doibase 10.1007/BF01298293} {\bibfield  {journal} {\bibinfo  {journal} {Zeitschrift f{\"u}r Physik B Condensed Matter}\ }\textbf {\bibinfo {volume} {42}},\ \bibinfo {pages} {65} (\bibinfo {year} {1981})}\BibitemShut {NoStop}%
\bibitem [{\citenamefont {Reeve}\ and\ \citenamefont {Guttmann}(1981)}]{boundary4}%
  \BibitemOpen
  \bibfield  {author} {\bibinfo {author} {\bibfnamefont {J.}~\bibnamefont {Reeve}}\ and\ \bibinfo {author} {\bibfnamefont {A.~J.}\ \bibnamefont {Guttmann}},\ }\href {\doibase 10.1088/0305-4470/14/12/028} {\bibfield  {journal} {\bibinfo  {journal} {Journal of Physics A: Mathematical and General}\ }\textbf {\bibinfo {volume} {14}},\ \bibinfo {pages} {3357} (\bibinfo {year} {1981})}\BibitemShut {NoStop}%
\bibitem [{\citenamefont {Diehl}(1997)}]{boundary5}%
  \BibitemOpen
  \bibfield  {author} {\bibinfo {author} {\bibfnamefont {H.~W.}\ \bibnamefont {Diehl}},\ }\href {\doibase 10.1142/S0217979297001751} {\bibfield  {journal} {\bibinfo  {journal} {International Journal of Modern Physics B}\ }\textbf {\bibinfo {volume} {11}},\ \bibinfo {pages} {3503} (\bibinfo {year} {1997})},\ \Eprint {http://arxiv.org/abs/https://doi.org/10.1142/S0217979297001751} {https://doi.org/10.1142/S0217979297001751} \BibitemShut {NoStop}%
\bibitem [{\citenamefont {Peskin}(1978)}]{peskindual}%
  \BibitemOpen
  \bibfield  {author} {\bibinfo {author} {\bibfnamefont {M.~E.}\ \bibnamefont {Peskin}},\ }\href {\doibase https://doi.org/10.1016/0003-4916(78)90252-X} {\bibfield  {journal} {\bibinfo  {journal} {Annals of Physics}\ }\textbf {\bibinfo {volume} {113}},\ \bibinfo {pages} {122 } (\bibinfo {year} {1978})}\BibitemShut {NoStop}%
\bibitem [{\citenamefont {Dasgupta}\ and\ \citenamefont {Halperin}(1981)}]{halperindual}%
  \BibitemOpen
  \bibfield  {author} {\bibinfo {author} {\bibfnamefont {C.}~\bibnamefont {Dasgupta}}\ and\ \bibinfo {author} {\bibfnamefont {B.~I.}\ \bibnamefont {Halperin}},\ }\href {\doibase 10.1103/PhysRevLett.47.1556} {\bibfield  {journal} {\bibinfo  {journal} {Phys. Rev. Lett.}\ }\textbf {\bibinfo {volume} {47}},\ \bibinfo {pages} {1556} (\bibinfo {year} {1981})}\BibitemShut {NoStop}%
\bibitem [{\citenamefont {Fisher}\ and\ \citenamefont {Lee}(1989)}]{leedual}%
  \BibitemOpen
  \bibfield  {author} {\bibinfo {author} {\bibfnamefont {M.~P.~A.}\ \bibnamefont {Fisher}}\ and\ \bibinfo {author} {\bibfnamefont {D.~H.}\ \bibnamefont {Lee}},\ }\href {\doibase 10.1103/PhysRevB.39.2756} {\bibfield  {journal} {\bibinfo  {journal} {Phys. Rev. B}\ }\textbf {\bibinfo {volume} {39}},\ \bibinfo {pages} {2756} (\bibinfo {year} {1989})}\BibitemShut {NoStop}%
\bibitem [{\citenamefont {Senthil}\ and\ \citenamefont {Fisher}(2005)}]{senthilfisher}%
  \BibitemOpen
  \bibfield  {author} {\bibinfo {author} {\bibfnamefont {T.}~\bibnamefont {Senthil}}\ and\ \bibinfo {author} {\bibfnamefont {M.~P.~A.}\ \bibnamefont {Fisher}},\ }\href@noop {} {\bibfield  {journal} {\bibinfo  {journal} {Phys. Rev. B}\ }\textbf {\bibinfo {volume} {74}},\ \bibinfo {pages} {064405} (\bibinfo {year} {2005})}\BibitemShut {NoStop}%
\bibitem [{\citenamefont {You}\ \emph {et~al.}(2014)\citenamefont {You}, \citenamefont {Bi}, \citenamefont {Rasmussen}, \citenamefont {Slagle},\ and\ \citenamefont {Xu}}]{YouXu2013}%
  \BibitemOpen
  \bibfield  {author} {\bibinfo {author} {\bibfnamefont {Y.-Z.}\ \bibnamefont {You}}, \bibinfo {author} {\bibfnamefont {Z.}~\bibnamefont {Bi}}, \bibinfo {author} {\bibfnamefont {A.}~\bibnamefont {Rasmussen}}, \bibinfo {author} {\bibfnamefont {K.}~\bibnamefont {Slagle}}, \ and\ \bibinfo {author} {\bibfnamefont {C.}~\bibnamefont {Xu}},\ }\href {\doibase 10.1103/PhysRevLett.112.247202} {\bibfield  {journal} {\bibinfo  {journal} {Phys. Rev. Lett.}\ }\textbf {\bibinfo {volume} {112}},\ \bibinfo {pages} {247202} (\bibinfo {year} {2014})}\BibitemShut {NoStop}%
\bibitem [{\citenamefont {Lee}\ \emph {et~al.}(2022)\citenamefont {Lee}, \citenamefont {You},\ and\ \citenamefont {Xu}}]{sptdecohere}%
  \BibitemOpen
  \bibfield  {author} {\bibinfo {author} {\bibfnamefont {J.~Y.}\ \bibnamefont {Lee}}, \bibinfo {author} {\bibfnamefont {Y.-Z.}\ \bibnamefont {You}}, \ and\ \bibinfo {author} {\bibfnamefont {C.}~\bibnamefont {Xu}},\ }\href {\doibase 10.48550/ARXIV.2210.16323} {\enquote {\bibinfo {title} {Symmetry protected topological phases under decoherence},}\ } (\bibinfo {year} {2022})\BibitemShut {NoStop}%
\bibitem [{\citenamefont {Zhang}\ \emph {et~al.}(2024)\citenamefont {Zhang}, \citenamefont {Qi},\ and\ \citenamefont {Bi}}]{zhang2024strange}%
  \BibitemOpen
  \bibfield  {author} {\bibinfo {author} {\bibfnamefont {J.-H.}\ \bibnamefont {Zhang}}, \bibinfo {author} {\bibfnamefont {Y.}~\bibnamefont {Qi}}, \ and\ \bibinfo {author} {\bibfnamefont {Z.}~\bibnamefont {Bi}},\ }\href@noop {} {\enquote {\bibinfo {title} {Strange correlation function for average symmetry-protected topological phases},}\ } (\bibinfo {year} {2024}),\ \Eprint {http://arxiv.org/abs/2210.17485} {arXiv:2210.17485 [cond-mat.str-el]} \BibitemShut {NoStop}%
\bibitem [{\citenamefont {Ma}\ \emph {et~al.}(2024)\citenamefont {Ma}, \citenamefont {Zhang}, \citenamefont {Bi}, \citenamefont {Cheng},\ and\ \citenamefont {Wang}}]{ma2024topological}%
  \BibitemOpen
  \bibfield  {author} {\bibinfo {author} {\bibfnamefont {R.}~\bibnamefont {Ma}}, \bibinfo {author} {\bibfnamefont {J.-H.}\ \bibnamefont {Zhang}}, \bibinfo {author} {\bibfnamefont {Z.}~\bibnamefont {Bi}}, \bibinfo {author} {\bibfnamefont {M.}~\bibnamefont {Cheng}}, \ and\ \bibinfo {author} {\bibfnamefont {C.}~\bibnamefont {Wang}},\ }\href@noop {} {\enquote {\bibinfo {title} {Topological phases with average symmetries: the decohered, the disordered, and the intrinsic},}\ } (\bibinfo {year} {2024}),\ \Eprint {http://arxiv.org/abs/2305.16399} {arXiv:2305.16399 [cond-mat.str-el]} \BibitemShut {NoStop}%
\bibitem [{\citenamefont {Garratt}\ \emph {et~al.}(2023)\citenamefont {Garratt}, \citenamefont {Weinstein},\ and\ \citenamefont {Altman}}]{altman1}%
  \BibitemOpen
  \bibfield  {author} {\bibinfo {author} {\bibfnamefont {S.~J.}\ \bibnamefont {Garratt}}, \bibinfo {author} {\bibfnamefont {Z.}~\bibnamefont {Weinstein}}, \ and\ \bibinfo {author} {\bibfnamefont {E.}~\bibnamefont {Altman}},\ }\href {\doibase 10.1103/PhysRevX.13.021026} {\bibfield  {journal} {\bibinfo  {journal} {Phys. Rev. X}\ }\textbf {\bibinfo {volume} {13}},\ \bibinfo {pages} {021026} (\bibinfo {year} {2023})}\BibitemShut {NoStop}%
\bibitem [{\citenamefont {Lee}\ \emph {et~al.}(2023)\citenamefont {Lee}, \citenamefont {Jian},\ and\ \citenamefont {Xu}}]{wfdecohere}%
  \BibitemOpen
  \bibfield  {author} {\bibinfo {author} {\bibfnamefont {J.~Y.}\ \bibnamefont {Lee}}, \bibinfo {author} {\bibfnamefont {C.-M.}\ \bibnamefont {Jian}}, \ and\ \bibinfo {author} {\bibfnamefont {C.}~\bibnamefont {Xu}},\ }\href {\doibase 10.1103/PRXQuantum.4.030317} {\bibfield  {journal} {\bibinfo  {journal} {PRX Quantum}\ }\textbf {\bibinfo {volume} {4}},\ \bibinfo {pages} {030317} (\bibinfo {year} {2023})}\BibitemShut {NoStop}%
\bibitem [{\citenamefont {Bao}\ \emph {et~al.}(2023)\citenamefont {Bao}, \citenamefont {Fan}, \citenamefont {Vishwanath},\ and\ \citenamefont {Altman}}]{altman2}%
  \BibitemOpen
  \bibfield  {author} {\bibinfo {author} {\bibfnamefont {Y.}~\bibnamefont {Bao}}, \bibinfo {author} {\bibfnamefont {R.}~\bibnamefont {Fan}}, \bibinfo {author} {\bibfnamefont {A.}~\bibnamefont {Vishwanath}}, \ and\ \bibinfo {author} {\bibfnamefont {E.}~\bibnamefont {Altman}},\ }\href@noop {} {\enquote {\bibinfo {title} {Mixed-state topological order and the errorfield double formulation of decoherence-induced transitions},}\ } (\bibinfo {year} {2023}),\ \Eprint {http://arxiv.org/abs/2301.05687} {arXiv:2301.05687 [quant-ph]} \BibitemShut {NoStop}%
\bibitem [{\citenamefont {Fan}\ \emph {et~al.}(2023)\citenamefont {Fan}, \citenamefont {Bao}, \citenamefont {Altman},\ and\ \citenamefont {Vishwanath}}]{fan2023}%
  \BibitemOpen
  \bibfield  {author} {\bibinfo {author} {\bibfnamefont {R.}~\bibnamefont {Fan}}, \bibinfo {author} {\bibfnamefont {Y.}~\bibnamefont {Bao}}, \bibinfo {author} {\bibfnamefont {E.}~\bibnamefont {Altman}}, \ and\ \bibinfo {author} {\bibfnamefont {A.}~\bibnamefont {Vishwanath}},\ }\href@noop {} {\enquote {\bibinfo {title} {Diagnostics of mixed-state topological order and breakdown of quantum memory},}\ } (\bibinfo {year} {2023}),\ \Eprint {http://arxiv.org/abs/2301.05689} {arXiv:2301.05689 [quant-ph]} \BibitemShut {NoStop}%
\bibitem [{\citenamefont {Yang}\ \emph {et~al.}(2023)\citenamefont {Yang}, \citenamefont {Mao},\ and\ \citenamefont {Jian}}]{jianmeasure2}%
  \BibitemOpen
  \bibfield  {author} {\bibinfo {author} {\bibfnamefont {Z.}~\bibnamefont {Yang}}, \bibinfo {author} {\bibfnamefont {D.}~\bibnamefont {Mao}}, \ and\ \bibinfo {author} {\bibfnamefont {C.-M.}\ \bibnamefont {Jian}},\ }\href {\doibase 10.1103/PhysRevB.108.165120} {\bibfield  {journal} {\bibinfo  {journal} {Phys. Rev. B}\ }\textbf {\bibinfo {volume} {108}},\ \bibinfo {pages} {165120} (\bibinfo {year} {2023})}\BibitemShut {NoStop}%
\bibitem [{\citenamefont {Zou}\ \emph {et~al.}(2023)\citenamefont {Zou}, \citenamefont {Sang},\ and\ \citenamefont {Hsieh}}]{zoumeasure}%
  \BibitemOpen
  \bibfield  {author} {\bibinfo {author} {\bibfnamefont {Y.}~\bibnamefont {Zou}}, \bibinfo {author} {\bibfnamefont {S.}~\bibnamefont {Sang}}, \ and\ \bibinfo {author} {\bibfnamefont {T.~H.}\ \bibnamefont {Hsieh}},\ }\href {\doibase 10.1103/PhysRevLett.130.250403} {\bibfield  {journal} {\bibinfo  {journal} {Phys. Rev. Lett.}\ }\textbf {\bibinfo {volume} {130}},\ \bibinfo {pages} {250403} (\bibinfo {year} {2023})}\BibitemShut {NoStop}%
\bibitem [{\citenamefont {Su}\ \emph {et~al.}(2024)\citenamefont {Su}, \citenamefont {Myerson-Jain},\ and\ \citenamefont {Xu}}]{cherndecohere}%
  \BibitemOpen
  \bibfield  {author} {\bibinfo {author} {\bibfnamefont {K.}~\bibnamefont {Su}}, \bibinfo {author} {\bibfnamefont {N.}~\bibnamefont {Myerson-Jain}}, \ and\ \bibinfo {author} {\bibfnamefont {C.}~\bibnamefont {Xu}},\ }\href {\doibase 10.1103/PhysRevB.109.035146} {\bibfield  {journal} {\bibinfo  {journal} {Phys. Rev. B}\ }\textbf {\bibinfo {volume} {109}},\ \bibinfo {pages} {035146} (\bibinfo {year} {2024})}\BibitemShut {NoStop}%
\bibitem [{\citenamefont {Myerson-Jain}\ \emph {et~al.}(2023)\citenamefont {Myerson-Jain}, \citenamefont {Hughes},\ and\ \citenamefont {Xu}}]{anyondecohere}%
  \BibitemOpen
  \bibfield  {author} {\bibinfo {author} {\bibfnamefont {N.}~\bibnamefont {Myerson-Jain}}, \bibinfo {author} {\bibfnamefont {T.~L.}\ \bibnamefont {Hughes}}, \ and\ \bibinfo {author} {\bibfnamefont {C.}~\bibnamefont {Xu}},\ }\href@noop {} {\enquote {\bibinfo {title} {Decoherence through ancilla anyon reservoirs},}\ } (\bibinfo {year} {2023}),\ \Eprint {http://arxiv.org/abs/2312.04638} {arXiv:2312.04638 [cond-mat.str-el]} \BibitemShut {NoStop}%
\bibitem [{\citenamefont {Toldin}\ \emph {et~al.}()\citenamefont {Toldin}, \citenamefont {Metlitski},\ and\ \citenamefont {Assaad}}]{toldinfuture}%
  \BibitemOpen
  \bibfield  {author} {\bibinfo {author} {\bibfnamefont {F.~P.}\ \bibnamefont {Toldin}}, \bibinfo {author} {\bibfnamefont {M.~A.}\ \bibnamefont {Metlitski}}, \ and\ \bibinfo {author} {\bibfnamefont {F.~F.}\ \bibnamefont {Assaad}},\ }\href@noop {} {}\Eprint {http://arxiv.org/abs/to appear} {to appear} \BibitemShut {NoStop}%
\bibitem [{\citenamefont {Dey}\ \emph {et~al.}(2020)\citenamefont {Dey}, \citenamefont {Hansen},\ and\ \citenamefont {Shpot}}]{BCFTNormalBCOPE}%
  \BibitemOpen
  \bibfield  {author} {\bibinfo {author} {\bibfnamefont {P.}~\bibnamefont {Dey}}, \bibinfo {author} {\bibfnamefont {T.}~\bibnamefont {Hansen}}, \ and\ \bibinfo {author} {\bibfnamefont {M.}~\bibnamefont {Shpot}},\ }\href {\doibase https://doi.org/10.1007/JHEP12(2020)051} {\bibfield  {journal} {\bibinfo  {journal} {J. High Energy Phys}\ }\textbf {\bibinfo {volume} {51}} (\bibinfo {year} {2020}),\ https://doi.org/10.1007/JHEP12(2020)051}\BibitemShut {NoStop}%
\end{thebibliography}%

\onecolumngrid
\appendix

\section{Review of the Extraordinary-Log boundary of the $3d$ $\O(N)$ model }

In this section, we will review the novel extraordinary-log boundary first proposed for the critical $\O(N)$ model (or $O(N)$ Wilson-Fisher) in three-dimensions in Ref.~\cite{maxboundary}. The critical $\O(N)$ model with boundary may be studied in a two-layer approach. The interactions coupling the boundary (layer-1) to the rest of the bulk (layer-2) are treated perturbatively. In the case that the boundary has strong tendency to order, this theory can be represented as a $2d$ non-linear sigma model with $\O(N)$-vector $\vec{n}$ living on the boundary which couples to the critical bulk $\O(N)$ fields $\vec{\phi}$ which see the ordinary boundary condition. However in this two-layer construction, as pointed out in Ref.~\cite{maxboundary}, it is more convenient to study the following effective IR theory instead if one assumes that the boundary prefers to order in the $N$-th direction
\begin{equation}
    \mathcal{S} = \mathcal{S}_\text{normal} - s \int \limits_{y_\perp = 0}  \dd[2]{x} \vec{t}(x) \cdot \vec{\pi}(x) + \frac{1}{2g} \int \limits_{y_\perp = 0} \dd[2]{x}  \bigg \{(\partial_\mu \vec{n})^2 - 2\vec{h} \cdot \vec{n} \bigg \} + \dots \label{SLogBoundaryIR}
\end{equation}
We denote the coordinate $y_\perp$ as the coordinate normal to the boundary. In Eq.~\ref{SLogBoundaryIR}, the boundary is placed at $y_\perp = 0$. In this expression, $S_\text{normal}$ is the critical bulk with normal boundary conditions, where the bulk $\O(N)$-vector $\vec{\phi}$ is chosen to point in the $N$-th direction at the boundary. The $\O(N)$-vector at the boundary is parameterized as $\vec{n} = (\vec{\pi},\sqrt{1 - \vec{\pi}^2})$ and $\vec{\pi}$ couples to the tilt operator $\vec{t}$, which is the boundary operator for the bulk fields $\phi_1, \dots, \phi_{N-1}$ at the normal boundary. This tilt operator has (protected) scaling dimension $\Delta_t = 2$ in three-dimensions. Lastly, the coefficient $s$ is fixed by symmetry to be a specific function of $N$, and hence is not allowed to flow under RG. 
At the normal boundary condition, the bulk fields obey the following bulk-boundary operator product expansion (OPE) near $y_\perp = 0$:

\begin{align}
    \phi_N(x,y_\perp) &\sim \frac{a}{(2 y_\perp)^{2 \Delta_\phi}} + \dots,\\
    \phi_{i = 1, \dots, N-1}(x,y_\perp) &\sim  \frac{b  }{(2y_\perp)^{ \Delta_\phi - 2}}t_i(x) + \dots,
\end{align}
where we've only kept the most relevant boundary operators in the bulk-boundary OPE (the identity is the most relevant operator in the expansion for $\phi_N$). The OPE coefficients $a$ and $b$ are universal quantities of the normal boundary condition and are known from a $4 - \epsilon$ expansion of the $\O(N)$ model \cite{BCFTNormalBCOPE}. A consequence of the $\O(N)$ symmetry of the theory is that the coefficient $s$ in Eq.~\ref{SLogBoundaryIR} is fixed in terms of these universal coefficients (and hence does not flow) to be $s = \frac{a}{b}$. Generically, as long as $a$ and hence $s$ is non-singular one should entertain the possibility of an extraordinary-log boundary condition. 

The extraordinary-log boundary phase in this theory manifests itself in the long-wavelength behavior of correlations functions of $\vec{n}(x)$ that decay logarithmically. To see this, we will review the calculation of the renormalized two-point function $\langle \vec{n}(x) \cdot \vec{n}(0) \rangle$ in the theory Eq.~\ref{SLogBoundaryIR}. It is convenient to renormalize the non-linear sigma model with counterterms in the MS-bar scheme. We will use the convention that the dimension of the boundary is lowered to $d = 2(1 - \epsilon)$ in dimensional regularization and the following definitions of bare couplings and fields (distinguished by a $0$-subscript) in terms of their renormalized counterparts:

\begin{equation}
    g_0 = \mu^{2 \epsilon} Z_g g, \text{ } n^{i = 1, \dots, N-1}_0 = \sqrt{Z_n} n^{i = 1, \dots, N-1}, \text{ and } \vec{h}_0 = \frac{Z_g}{\sqrt{Z_n}} \vec{h} \label{renorm1}.
\end{equation}
The factor $\vec{h}= (0,h)$ is merely an IR regulator that picks out the $N$-th direction, and $\mu$ is the running energy scale. It is easier to work with the theory perturbatively in terms of the $\Vec{\pi}$-fields, which are renormalized by $\vec{n}_0 = (\sqrt{Z_n} \vec{\pi}, \sqrt{1 - Z_n \vec{\pi}^2})$. Furthermore, we will choose to rescale $\vec{\pi} \rightarrow \sqrt{g} \vec{\pi}$. After this, the theory can be written to leading order in the coupling $g$ as

\begin{align}
    \mathcal{S}_\text{NLSM} =  \frac{\mu^{-2 \epsilon}}{2} \int \dd[d]{x} Z_\pi (\partial_\mu \vec{\pi})^2 + Z_m h \vec{\pi}^2 + g Z_{g}' (\vec{\pi} \cdot \partial_\mu \vec{\pi})^2 - \frac{gh Z_g' Z_m Z_{\pi}^{-1}}{4} (\vec{\pi}^2)^2 + \dots
\end{align}
In terms of the renormalization of the coupling and the original boundary $\O(N)$-vector in Eq.~\ref{renorm1}, the factors $Z_\pi$ and $Z_m$, which renormalize the $\vec{\pi}$-field propagator, and $Z_g'$ are $Z_n = Z_m^2$, $Z_g = Z_g' Z_\pi^2$. As there is only renormalization of $\vec{n}$ and $g$ in the original theory, these factors obey the constraint $Z_g' = Z_\pi Z_m^2$. A standard calculation of the $\vec{\pi}$-field self-energy at 1-loop shows that in the non-linear sigma model to leading order, one requires at $\mathcal{O}(g)$

\begin{equation}
    Z_\pi = 1 - \frac{g}{4 \pi \epsilon} \text{ and } Z_m = 1 - \frac{g(N-1)}{8 \pi \epsilon} \Longrightarrow 
    \begin{cases}
        Z_g = 1 - \frac{g(N-2)}{4 \pi \epsilon},\\
        Z_n = 1 - \frac{g(N-1)}{4 \pi \epsilon}.
    \end{cases}
\end{equation}
From these factors, the $\beta$-function for the coupling $g$ and the anomalous dimension of the $\O(N)$-vector can be computed as
\begin{align}
\beta(g) &= \dv{g}{\ln \mu} \simeq -\frac{g^2(N-2)}{4 \pi}
,\\
    \gamma_n &= \frac{1}{2} \dv{\ln Z_n}{\ln \mu} \simeq \frac{g(N-1)}{4 \pi}.
\end{align}
These are merely the well-known 1-loop results for the non-linear sigma model. With the inclusion of the coupling to the bulk through the tilt operator, the coupling $g$ is further renormalized. The appropriately regularized leading order correction to self-energy of the $\vec{\pi}$-field due to the tilt operator is
\begin{equation}
    \delta \Pi^{ij}(q) = - g s^2 \delta^{ij} \int \dd[2]{r} \frac{e^{iq \cdot r}-1}{|r|^4} \simeq \delta^{ij}\frac{\pi^2 s^2 q^2}{4 \pi \epsilon}  + \text{finite}
\end{equation}
Crucially, requiring that there is an $\O(N)$ symmetry at low energies once the IR regulator is taken to zero implies that there cannot be a mass for the $\vec{\pi}$ generated through renormalization that is independent of $h$. The consequence of this is that the correction from coupling to the bulk via the tilt operator only changes $Z_\pi$ and not $Z_m$. The factor $Z_\pi$ now becomes
\begin{equation}
    Z_\pi = 1- \frac{g(1+\pi^2 s^2)}{4 \pi \epsilon} \Longrightarrow Z_g = 1-\frac{g(N-2 - \pi^2 s^2)}{4 \pi \epsilon}.
\end{equation}
As only $Z_\pi$ is changed through the coupling to the bulk, the $\beta$-function will gain a new term while the anomalous dimension of $\vec{n}$ remains intact. In particular, the $\beta$-function is now (with solution)
\begin{equation}
    \beta(g) = - \frac{g^2(N - 2 - \pi^2 s^2)}{4 \pi} \Longrightarrow g(\mu) = \frac{g_0}{1 + \frac{g_0(N - 2 - \pi^2 s^2)}{2 \pi} \log(\mu / \mu_0)}.
\end{equation}
For $N = 2$, which corresponds to the case of bulk $\U(1)$ Wilson-Fisher/$3$D XY transition as discussed in the main text, the coupling is marginally irrelevant and flows slowly back to the $g= 0$ fixed point. Along this RG flow, the renormalized two-point function $\langle \vec{n}(x) \cdot \vec{n}(0) \rangle$ can be computed by solving the Callan-Symanzik equation when $h_0 = 0$, 
\begin{align}
    \bigg ( \mu \pdv{\mu} + \beta(g) \pdv{g} + 2 \gamma_n \bigg ) \langle \vec{n}(x) \cdot \vec{n}(0) \rangle =0 \Longrightarrow \langle \vec{n}(x) \cdot \vec{n}(0) \rangle &\sim \exp \bigg (-2 \int_0^{\log(\mu |x|)} \dd{\log (y)} \gamma(g(y))   \bigg ),\\
    &\sim \frac{1}{(\log(\mu |x|))^{ -(\frac{N-1}{N-2 - \pi^2 s^2})}}.
\end{align} We can see that as long as $s$ is nonzero, the extraordinary-log solution always exists for $N = 2$. Hence we expect the extraordinary-log boundary to exist generally for an XY order parameter, even if the QCP in the bulk involves fractionalization. 

\section{Derivation of super power-law correlation}

In this appendix, we present a derivation of the two-point function Eq.~\ref{green0}. Following Ref.~\cite{altman1}, we employ the wave-functional method to derive the equal-time correlation $\langle e^{\mathtt{i}\phi(x)}e^{-\mathtt{i}\phi(0)}\rangle$ for the dual boson $\phi$. By definition, the wave functional is given by 
\begin{flalign}
|\Psi\rangle=\int\mathcal{D}[\theta]e^{-\mathcal{H}[\theta]}|\theta\rangle,
\end{flalign}
which satisfies $\langle\Psi|\theta(x)\theta(0)|\Psi\rangle=G_{\theta}(\tau=0,x)$, and $\mathcal{H}[\theta]=\int\frac{\textrm{d}k}{2\pi}G_{\theta}^{-1}(\tau=0,k)|\theta(k)|^{2}$ at the Gaussian level. 

Within the Gaussian approximation, the extraordinary-log correlation $\langle e^{\mathtt{i}\theta(\mathbf{x})}e^{-\mathtt{i}\theta(0)}\rangle\sim1/(\log(\mu|\mathbf{x}|))^{q/4}$ implies $G_{\theta}(\mathbf{x})=-(q/4)\log(\log(\mu|\mathbf{x}|))$, where $\mathbf{x}=(\tau,x)$ denotes the boundary coordinate. We begin with the Fourier transformation of $G_{\theta}(\mathbf{x})$ under the long-wavelength limit. First, we expand $G_{\theta}(\mathbf{x})$ in terms of its Taylor series
\begin{flalign}
G_{\theta}(\mathbf{x})=-\frac{q}{4}\left(\log(\log\mu)+\sum_{n=1}^{+\infty}\frac{(-1)^{n+1}}{n}\left(\frac{\log|\mathbf{x}|}{\log\mu}\right)^{n}\right).
\end{flalign}
Next, we extract the most singular contribution of the Fourier transformation for each term. Let us perform the integral in general dimensions 
\begin{flalign}
\int\textrm{d}^{d}\mathbf{x}\frac{(-1)^{n+1}}{n}\left(\frac{\log|\mathbf{x}|}{\log\mu}\right)^{n}e^{\mathtt{i}\mathbf{k}\cdot\mathbf{x}}&=\int_{0}^{+\infty}\textrm{d}x\frac{(-1)^{n+1}}{n}\left(\frac{\log x}{\log\mu}\right)^{n}(2\pi)^{\frac{d}{2}}x^{\frac{d}{2}}|\mathbf{k}|^{1-\frac{d}{2}}\mathsf{BesselJ}\left[\frac{d-2}{2},x|\mathbf{k}|\right]\nonumber\\&=-\frac{(2\pi)^{d}}{\Omega_{d-1}}\frac{(\log|\mathbf{k}|)^{n-1}}{|\mathbf{k}|^{d}(\log\mu)^{n}}+(\textrm{less singular terms under }|\mathbf{k}|\rightarrow0),
\end{flalign}
where $\Omega_{d-1}=2\pi^{\frac{d}{2}}/\Gamma(\frac{d}{2})$ represents the volume of $(d-1)$-sphere. The resummation of the Taylor series leads to
\begin{flalign}
G_{\theta}(\mathbf{k})=\frac{q}{4}\sum_{n=1}^{+\infty}\frac{(2\pi)^{d}}{\Omega_{d-1}}\frac{(\log|\mathbf{k}|)^{n-1}}{|\mathbf{k}|^{d}(\log\mu)^{n}}+\ldots=-\frac{q}{4}\frac{(2\pi)^{d}}{\Omega_{d-1}}\frac{1}{|\mathbf{k}|^{d}\log(|\mathbf{k}|/\mu)}+\ldots
\end{flalign}
The Fourier transformation of the equal-time correlation $G_{\theta}(\tau=0,x)=-(q/4)\log(\log(\mu|x|))$ can be readily obtained by setting $d=1$ in the above expression
\begin{flalign}
G_{\theta}(\tau=0,k)=-\frac{q}{4}\frac{\pi}{|k|\log(|k|/\mu)}+\ldots
\label{2pt_theta}
\end{flalign}
Going to the dual basis $|\phi\rangle$, the wave functional becomes 
\begin{flalign}
|\Psi\rangle=\int\mathcal{D}[\phi]\mathcal{D}[\theta]\exp\left(-\mathcal{H}[\theta]+\int\textrm{d}x\frac{\mathtt{i}}{\pi}\phi\partial_{x}\theta\right)|\phi\rangle,
\end{flalign}
where we utilize $\langle\phi|\theta\rangle=\exp\left(\frac{\mathtt{i}}{\pi}\int\textrm{d}x\phi\partial_{x}\theta\right)$, originating from the Wess-Zumino-Witten term of the Luttinger liquid. Now, we are prepared to compute the equal-time Green function of $\phi$
\begin{flalign}
G_{\phi}(\tau=0,x)=\langle\Psi|\phi(x)\phi(0)|\Psi\rangle=\int\frac{\textrm{d}k}{2\pi}\frac{\pi^{2}}{k^{2}}G_{\theta}^{-1}(\tau=0,k)e^{\mathtt{i}kx}=-\frac{2}{q}(\log(\mu|x|))^{2}+\ldots
\end{flalign}
Within the Gaussian approximation, we find $\langle e^{\mathtt{i}\phi(x)}e^{-\mathtt{i}\phi(0)}\rangle\sim e^{-(2/q)(\log|x|)^{2}}$. Consequently, we can determine the Green function Eq.~\ref{green0} of the fermion modes accordingly.

\end{document}